\documentclass{emulateapj}
\usepackage{amsmath} 
\usepackage{graphicx}

\newcommand{\myemail}{nathaniel.roth@berkeley.edu}

\shorttitle{Radiation Pressure on the Dusty ``Torus''}
\shortauthors{Roth et al.}

\begin{document}

\title{3-D Radiative Transfer Calculations of Radiation Feedback from Massive Black Holes: Outflow of Mass from the Dusty ``Torus''}
   
\author{Nathaniel Roth\altaffilmark{1}}
\author{Daniel Kasen\altaffilmark{1,2,3}}
\author{Philip F. Hopkins\altaffilmark{2}} 
\author{Eliot Quataert\altaffilmark{1,2}}
\email{\myemail}

\altaffiltext{1} {Physics Department, University of California,
    Berkeley, CA 94720, USA}
\altaffiltext{2} {Astronomy Department and Theoretical Astrophysics Center, University of California,
    Berkeley, CA 94720, USA}
\altaffiltext{3} {Nuclear Science Division, Lawrence Berkeley National Laboratory, Berkeley, CA 94720, USA}
    
\begin{abstract}
Observational and theoretical arguments suggest that the momentum carried in mass outflows from AGN can reach several times $L / c$, corresponding to outflow rates of hundreds of solar masses per year. Radiation pressure on resonant absorption lines alone may not be sufficient to provide this momentum deposition, and the transfer of reprocessed IR radiation in dusty nuclear gas has been postulated to provide the extra enhancement. The efficacy of this mechanism, however, will be sensitive to multi-dimensional effects such as the tendency for the reprocessed radiation to preferentially escape along sightlines of lower column density. We use Monte Carlo radiative transfer calculations to determine the radiation force on dusty gas residing within approximately 30 parsecs from an accreting super-massive black hole. We calculate the net rate of momentum deposition in the surrounding gas and estimate the mass-loss rate in the resulting outflow as a function of solid angle for different black hole luminosities, sightline-averaged column densities, clumping parameters, and opening angles of the dusty gas. We find that these dust-driven winds carry momentum fluxes of 1-5 times $L/c$ and correspond to mass-loss rates of 10-100 $M_\odot$ per year for a $10^8$ $M_\odot$ black hole radiating at or near its Eddington limit. These results help to explain the origin of high velocity molecular and atomic outflows in local ULIRGs, and can inform numerical simulations of galaxy evolution including AGN feedback.
\end{abstract}

 
\section{Introduction}
\subsection{Motivations from observations and theory}

The nature of the interaction between an accreting super-massive black hole (SMBH) and its host galaxy remains a challenging problem in the study of galaxy evolution. Numerical simulations reveal that gas can be drawn inward toward the nucleus by gravitational torques arising from a series of gravitational instabilities \citep{Hopkins2010-1}. This gas typically forms a dusty structure at small radii with a characteristic length scale of $\sim$1-10 parsecs which in some cases has been imaged directly \citep{Jaffe2004, Raban2009}. Phenomenologically, this structure can be modelled as a torus \citep{Lawrence1991, Antonucci1993}, but there is an ongoing theoretical effort to provide a detailed, self-consistent explanation of its configuration and what supports it. If a sufficiently strong poloidal magnetic field is present at the parsec scale, one possible explanation is that the dusty gas is launched as a hydromagnetic wind \citep{Konigl1994, Keating2012}. Heating of the ISM from stellar feedback might support the dusty gas in a puffy disk \citep{Hopkins2012-1}. The disk might be simultaneously supported by infrared radiation pressure \citep{Pier1992-1, Krolik2007}, or the infrared radiation pressure may generate a failed wind \citep{Dorodnitsyn2011-1, Dorodnitsyn2012}.

Regardless of what supports the torus, gas continues to be drawn in to the black hole accretion disk at small radii ($< 10^{17}$ cm), where it powers an active galactic nucleus (AGN). The radiation emitted from SMBH accretion disks influences the dynamics of the torus itself, along with the dynamics of the host galaxy. This feedback may act through a number of channels that include radiative heating \citep[e.g.][]{DiMatteo2005}, jets \citep{Silk2005, Croton2006, McNamara2007}, and winds driven by radiation pressure on resonant ultraviolet lines \citep{Murray1995, Proga2000} and dust \citep{Konigl1994, Murray2005, Keating2012}. Our challenge is to understand the combined effect of all these modes of interaction. Improving our understanding of this connection will be crucial for answering questions about the growth of SMBHs, observations of AGN, and the star formation histories in galaxies.

Recent observations have begun to reveal the violent impact that AGN may have on their host galaxies. Observations of obscured quasars such as SDSS J1356+1026 have revealed outflows extending out to tens of kiloparsecs from the galactic nucleus \citep{Greene2012}. The estimated mechanical luminosity of these outflows ($10^{44-45}$ ergs s$^{-1}$) is too large to be explained by the inferred star formation activity. Other obscured quasars possess more massive outflows, with mass-loss rates of hundreds of solar masses per year \citep{Moe2009,Dunn2010}. Meanwhile, observations of local ultra-luminous infrared galaxies (ULIRGs) have led to the discovery of outflows with velocities that are correlated with the AGN bolometric luminosity \citep{Sturm2011}. These outflows also have mass-loss rates equal to several times the star formation rate and in some cases exceeding 1000 solar masses per year, depleting the gas on timescales as short as $10^6$ years. Adding to our picture are studies of post-starburst galaxies, exhibiting outflows with median velocity of approximately 1000 km s$^{-1}$, suggesting that past AGN activity played a role in launching the gas \citep{Tremonti2007}.

These observations are complemented by numerical simulations of AGN feedback \citep{Ciotti2010, DeBuhr2011-1, DeBuhr2011-2} that account for deposition of both energy and momentum from the accretion radiation, including a combination of heating by X-rays and photoionizations, radiation pressure at the kiloparsec scale, and winds driven from within a radius of less than 100 parsecs. Taken together, these effects can help to explain both the $M_\mathrm{BH}$ - $\sigma$ relation \citep{Ferrarese2000, Gebhardt2000,Tremaine2002} and the existence of galactic outflows observed at speeds of thousands of km s$^{-1}$. The results, particularly those of \citet{DeBuhr2011-2}, also suggest that line-driven winds may be insufficient to drive observed outflows, and that a large amount of momentum ($\gtrsim 3 \, L/c$) may need to be deposited via absorption by dust grains during the period when the SMBH is optically thick to both ultraviolet and far-infrared radiation, the time when most black hole growth is believed to occur \citep{Fabian1999, Hopkins2005-1}. 

A large uncertainty in the numerical calculations referenced above is the amount of radiative momentum deposited within the central unresolved radius. The velocity and mass-loss rate of the resulting wind depend sensitively on this coupling. Moreover, in those studies the momentum was deposited in a spherically symmetric manner. In reality, multidimensional effects, such as the tendency for radiation to escape out the rarefied, polar regions of the gas distribution, will be crucial. These effects have been considered by several previous studies. \citet{Pier1992-1} computed the radiation forces exerted on a torus modelled as a constant density cylindrical shell, and \citet{Krolik2007} extended that work to account for a more self-consistent rearrangement of the gas under the influence of the radiation. A radiation-hydrodynamics study that linked the effects of Compton scattering and broad absorption line winds at the parsec scale with inflow processes on galactic (kiloparsec) scales in two spatial dimensions was undertaken by \citet{Novak2011}, and this was extended in order to capture the radiative transfer through dusty gas in \citet{Novak2012}. Our study extends this work further by performing three dimensional Monte Carlo radiative transfer calculations for dusty gas, including both smooth and clumpy gas distributions, and by integrating the force on columns of gas in order to quantify the mass outflow rate from AGN radiating at high luminosity.

The momentum flux in radiation from a SMBH accretion disk with luminosity $L$ is $L/c$. Generally $L$ will not exceed $L_\mathrm{Edd}$, the Eddington luminosity set by the electron scattering (Thomson) opacity. Dust will contribute to the opacity seen by the radiation at large radii, but only at distances greater than the radius $r_\mathrm{sub}$ at which its temperature drops below the sublimation temperature $T_\mathrm{sub} \approx 1400 \; \mathrm{K}$. Although the sublimation temperature varies for each grain depending on its composition and its size, we choose to adopt the simplification of assigning a uniform sublimation temperature to all the dust in our calculations. The sublimation radius may be estimated as

\begin{align}
r_\mathrm{sub} &\approx \sqrt{\frac{L}{4 \pi \sigma_{\mathrm{SB}}T_{\mathrm{sub}}^4}}\nonumber \\
&= 0.62 \mathrm{\;pc} \left(\frac{L}{10^{46} \mathrm{\; erg s}^{-1}} \right)^{1/2}\left( \frac{T_\mathrm{sub}}{1400 \mathrm{\; K}}\right)^{-2} \; .
\end{align}
When the gas distribution surrounding the SMBH is not isotropic, $r_\mathrm{sub}$ may vary with angle. Within this radius, electron scattering dominates the opacity, and the usual Eddington limit applies. 

Once the intrinsic photons from the accretion disk encounter dust in the surrounding gas, they are absorbed and the energy is re-emitted at infrared wavelengths. If the gas is also optically thick to the infrared, then the re-emitted radiation will continue to be absorbed and re-emitted in a random-walk pattern until it exits the optically thick region. Along the way, momentum will be imparted by the photons to the gas multiple times. In this scenario it is possible for the radiation to transfer momentum to the gas at a rate that exceeds $L_\mathrm{Edd}/c$. For a spherically symmetric problem, this ``boost'' factor to the infrared radiation force is exactly the infrared optical depth of the gas, which can be shown as follows: In steady-state, when radiative equilibrium holds and the luminosity as a function of radius is constant, we may compute the radiation force per volume $f_\mathrm{rad}$ as 
\begin{equation}
f_\mathrm{rad} = \frac{L}{4 \pi \, r^2 \, c}\, \rho(r) \, \kappa(r) \; .
\end{equation}

The total outward force exerted by the radiation is
\begin{equation}
\int _V f_\mathrm{rad}\, dV = 4 \pi \left(\frac{L}{4 \pi \, c}\right) \int_{0}^{\infty} \rho(r) \, \kappa(r) \, dr = \tau \frac{L}{c}\; , 
\end{equation}
where $\tau$ is the radial optical depth for the infrared photons.

In a gas rich galactic nucleus with a column density of $10^{25}$ cm$^{-2}$, a mean mass per particle of 1.5 times the proton mass, and an infrared dust opacity of 10 cm$^{2}$ per gram of gas, an initial guess for the optical depth would be approximately 250. There are two primary effects that will reduce the actual radiation force from such a high value. The first is the lack of spherical symmetry: a torus obscures only a fraction of the solid angle surrounding the accretion disk, and the presence of clumps and voids in the torus can increase the photon mean free path for certain sightlines. The second effect is dust sublimation: dust will be absent from the innermost regions of the nucleus that contribute a substantial fraction to the gas column density, and the force integral can be well-approximated by setting its lower limit to $r_\mathrm{sub}$.

To get a sense of the sort of momentum deposition rates that have been observed, consider the case of Mrk 231. This system features an outflow of neutral gas with velocities in the range 360-900 km s$^{-1}$ and a mass-loss rate estimated at 420 solar masses per year \citep{Rupke2011}. The momentum flux in the outflow, estimated by multiplying the mass loss rate by the velocity, is between 2.6 to 6.5 times $L/c$ where $L$ is measured to be $1.1 \times 10^{46}$ ergs s$^{-1}$. The kinetic luminosity of the outflow, on the other hand, is estimated at $7.3 \times 10^{43}$ ergs s$^{-1}$, less than 1\% of the bolometric AGN luminosity.   

Modeling the force from radiation pressure, and predicting by what factor it exceeds $L/c$, becomes a difficult problem to tackle analytically in the absence of spherical symmetry, the presence of clumps, and with an accounting for dust sublimation. For these reasons, we turn here to three-dimensional radiative transfer calculations using the wavelength-dependent Monte Carlo radiative transfer code SEDONA \citep{Kasen2006}. Given that the radiative diffusion time in these systems is shorter than the dynamical times, we restrict ourselves to steady-state configurations that do not include an explicit coupling to hydrodynamics. 

In section \ref{MethodologySection}, we describe how we parametrize the gas configurations surrounding the black hole and how we treat the key physical processes in the radiative transfer. In section \ref{Results}, we present our results for a series of calculations in which we vary the opening angle of the torus, the amount of gas present, and the accretion disk luminosity. We also examine how our dynamical conclusions are affected by accounting for a clumpy rather than smooth distribution of dust and gas. Finally, in section \ref{Conclusion} we present our conclusions.

\section{Methodology}
\label{MethodologySection}
\subsection{Initial gas configuration - parameterized, smooth model}
\label{DiskGeometry}
Although the specific region we are studying is difficult to observe directly, gravito-hydrodynamic simulations \citep{Hopkins2012-1} provide information about its configuration before the effects of radiative feedback are felt. The gas and stars form a thick disk roughly in vertical hydrostatic equilibrium (our usage of the word ``disk'' throughout the remainder of this study refers to what is usually labelled as the torus and should not be confused with a reference to the black hole accretion disk, which is unresolved at our scales of interest). The puffiness of the disk in the \citet{Hopkins2012-1} simulations is to some extent determined by the sub-grids turbulent velocity dispersion when strong stellar feedback in the ISM is included, but also by bending modes (firehose instabilities) driven by resolved velocities when less stellar feedback is included. While further accretion of the gas at this scale will rely on non-axisymmetric torques, we first adopt a simple axisymmetric, hydrostatic disk model analogous to one used in \citet{Hopkins2012-1}. This parametrization captures the key features of the gas configuration seen in the hydrodynamics simulations, but allows us greater control over free parameters and removes unnecessary complications in our attempt to isolate the effects of the radiation. Such a parametrization also allows us to systematically introduce clumpiness into the gas for certain calculations (which, among other effects, breaks axisymmetry), as will be described in section \ref{ClumpySetup}

The vertical structure of the smooth disk model may be calculated by solving the equation of hydrostatic equilibrium in the normal ($z$) direction, assuming an isothermal equation of state with effective sound speed $c_s$ set by both the resolved and sub-grid velocity dispersion, along with any contribution from the thermal pressure of the gas, 
\begin{equation}
\frac{{c_s}^2}{\rho} \, \frac{d \rho }{dz} = -\frac{d \Phi}{dz} \; ,
\end{equation}
with solution 
\begin{equation}
\rho(R,z) = \rho(R,0) \exp \left\{c_s^{-2}\left[\Phi(R,0) - \Phi(R,z) \right] \right\} \; .
\end{equation} 
Here $\Phi$ denotes the gravitational potential, $\rho$ denotes the density of the gas, and $R$ is the cylindrical radius. If we assume that the gravitational potential is dominated by the mass of the central black hole $M_\mathrm{BH}$ at these scales, then the density distribution is 
\begin{equation}
\rho(R,z) = \rho(R,0) \exp \left\{ \frac{G M_\mathrm{BH} }{R \, c_s^2}\left[\frac{1}{\sqrt{1 + z^2/R^2}} - 1 \right] \right \}\;.
\end{equation}
In the limit of small $z/R$, this yields a Gaussian vertical structure. In this limit, the ratio of the squared sound speed to the squared Keplerian velocity $V_c$ functions as the ratio of the disk scale height to the cylindrical radius, and for convenience we choose to define a parameter that makes this identification universal:
\begin{equation}
\frac{h_s}{R} \equiv \frac{c_s}{V_c} = c_s \left( \frac{G\, M_\mathrm{BH}}{R} \right )^{-1/2} \;.
\end{equation}

Moderately large values of $h/R$ ($\gtrsim 0.2$) are suggested by the observed fraction of obscured versus unobscured quasars, although generally this fraction correlates strongly with luminosity \citep{Maiolino2007}. Meanwhile, typical values of $h_s/R$ found in \citet{Hopkins2012-1} range from 0.1 to 0.5. In this study we will consider $h_s/R$ in the range 0.1 to 0.35.

Mid-IR interferometric observations find the mid-plane density may be well-fit with a power-law $R^{-\gamma}$ where $\gamma$ lies within a range of approximately 0.4 to 1.4, with a tendency toward larger values for more luminous AGN \citep{Kishimoto2011}. This is also in agreement with the simulations presented in \citet{Hopkins2012-1} in which $\gamma \approx 1.5$. We find that the results for varying $\gamma$ correlate strongly with the corresponding change in torus mass within the computational domain, and so we choose to capture variations in torus mass and column density by varying the normalization of the radial density profile (the parameter $\rho_0$ described below), while fixing $\gamma$ to $1.5$ for all calculations presented in this paper. 

Converting to spherical polar coordinates $r$ and $\theta$, where $\theta$ is taken to be zero along the z-axis, we obtain 
\begin{equation}
\rho(r,\theta) = \rho_0 \left(\frac{r \sin \theta}{r_0}\right)^{-\gamma} \exp \left[(h_s/R)^{-2}(\sin{\theta} - 1) \right] \; .
\end{equation}
Here $r_0$ represents some inner cut-off radius where the density is $\rho_0$, and to prevent the radial column density from diverging we take $\rho(r < r_0) = \rho_0$.  

One undesirable aspect of this model is that it leads to an accumulation of mass in the polar region of the disk, where $\sin \theta$ is small. To correct for this, we allow the density profile to drop as a power law in the spherical radius $r$ rather than in the cylindrical radius $R$. This amounts to dropping the factor of $(\sin{\theta})^{-\gamma}$, which is only significant far from the disk mid-plane. This leaves
\begin{equation}
\rho(r,\theta) = \rho_0 \left(\frac{r }{r_0}\right)^{-\gamma} \exp \left[(h_s/R)^{-2}(\sin{\theta} - 1) \right] \;,
\end{equation}
which is quite similar to the phenomenological models of \citet{Granato1994} and \citet{Efstathiou1995} that were used to explain the properties of spectral energy distributions observed in dusty AGN.

The results from \citet{Hopkins2012-1} indicate that $h_s/R$ does not change by more than a factor of order unity for all $R$. For simplicity, we take $h_s/R$ to be a constant for all $R$ and allow it to vary as a free parameter for different disk models. For all calculations in this study we assume a black hole mass $M_\mathrm{BH}$ of $10^8$ $M_\odot$, and we parametrize the luminosity as a fraction of the electron-scattering Eddington luminosity for that mass. As mentioned above, we also vary $\rho_0$, which sets the sightline-averaged column density $\overline{N}_\mathrm{H}$. Unless stated otherwise, $\overline{N}_\mathrm{H}$ corresponds to the column density integrated to a distance of $r_0 = 0.1$ parsecs from the central black hole. Also, unless $\overline{N}_\mathrm{H}$ is being varied explicitly, $\rho_0$ is set so that the sightline-averaged column density is $3.4 \times 10^{24}$ cm$^{-2}$, with a mid-plane column density of $1.0 \times 10^{25}$ cm$^{-2}$. These values are consistent with the calculations from \citet{Hopkins2010-1} of surface densities of $10^{11}$ -- $10^{12}$ $M_\odot$ kpc$^{-2}$ for the central 10 parsecs surrounding the black hole. The fiducial parameters are summarized in Table \ref{FiducialParameters}. 

\begin{figure}

\begin{minipage}[b]{\textwidth}
\includegraphics[width=0.5\textwidth]{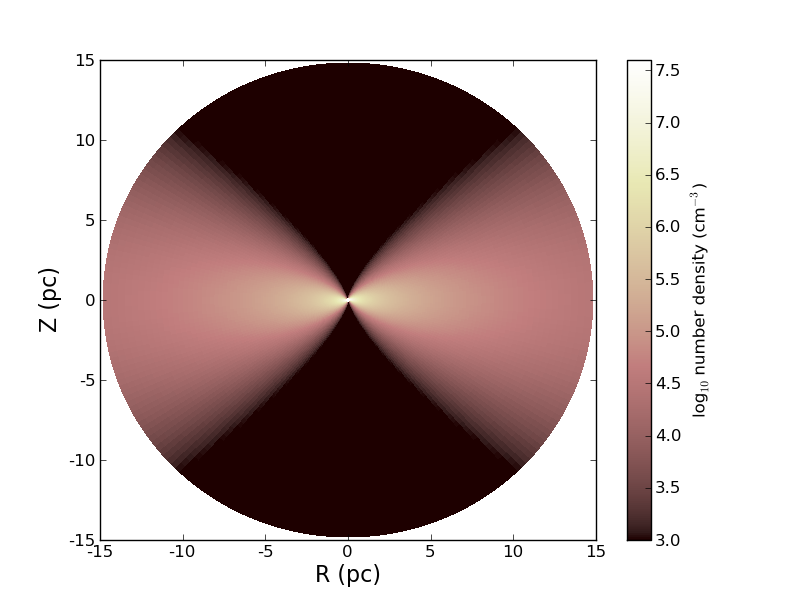}
\end{minipage}

\begin{minipage}[b]{\textwidth}
\includegraphics[width=0.5\textwidth]{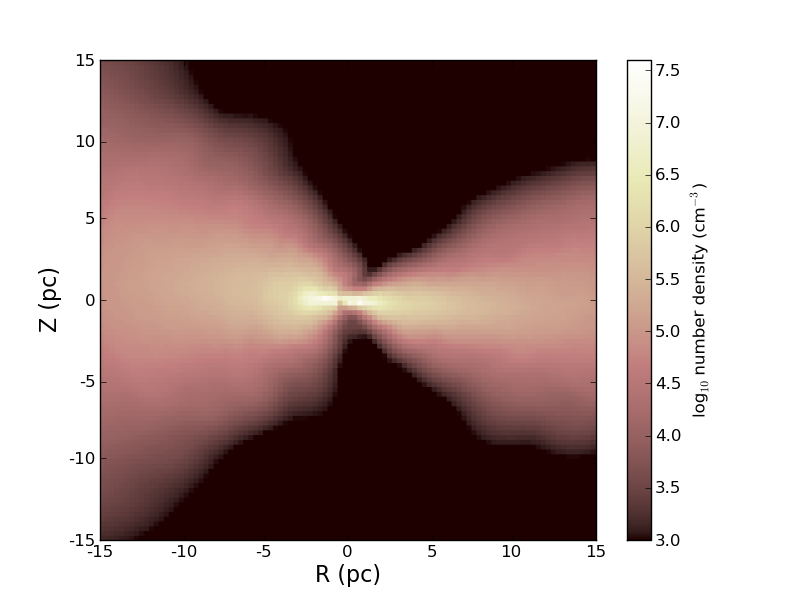}
\end{minipage}

\caption{Top: An example of a slice through the smooth model density distribution with the fiducial parameters listed in Table \ref{FiducialParameters}, except $\overline{N}_\mathrm{H} = 1.0 \times 10^{25}$ cm$^{-2}$ (chosen to match the simulation in the bottom panel). Bottom: A density slice taken from a hydrodynamical simulation of gas accretion onto a central black hole (see \citet{Hopkins2012-1}). Note that the color map is truncated at $n = 10^3$ cm$^{-2}$, and the density in the model distribution continues to drop below this value. }
\label{DensitySlices}
\end{figure}

For this smooth density model we use a two-dimensional grid with spherical polar $(r,\theta)$ coordinates, with logarithmic spacing in the radial coordinate and linear spacing in the angular coordinate. Our resolution is 192 radial zones and 64 $\theta$ zones for $\theta$ ranging from $0$ to $\pi/2$, with an assumed symmetry for $\theta \rightarrow \pi - \theta$. The radial zones span radii ranging from $r_0 = 0.1$ pc to an outer radius $r_\mathrm{out} =10^{20}$ cm ($\approx$ 32.4 pc). The 0.1 pc scale was chosen because it is a larger scale than the typical black hole accretion disk, but also small compared to the typical dust sublimation radius. We ignore all momentum deposition inside the 0.1 pc radius, and since nearly all the momentum deposition occurs at and beyond the sublimation radius, the exact choice of innermost radius has little effect on our results. Slices of the gas density for the model developed in this section, along with a simulation from \citet{Hopkins2012-1}, are shown in Figure \ref{DensitySlices}. 

\begin{deluxetable}{ccccc}
\tablecaption{Fiducial parameters}
\tablehead{\colhead{$h_s/R$} & \colhead{$\overline{N}_\mathrm{H}$ ( cm$^{-2}$ )} & \colhead{radial density} & \colhead{$L/L_\mathrm{Edd}$} & \colhead{$M_\mathrm{BH}$ ($M_\odot$)} \\ \colhead{} & \colhead{} & \colhead{power-law $\gamma$} & \colhead{} & \colhead{}}
\startdata
0.3 & $3.4 \times 10^{24}$ & 1.5 & 1 & $10^8$   \\
\enddata
\tablecomments{The first three parameters set the gas density distribution, while the last two set the relative strengths of the radiation pressure and gravity. The mean mass per particle is always set to $1.5$ times the proton mass throughout this paper. Note that the column density presented in this table corresponds to integrating the gas density from large radii to a distance of 0.1 pc from the BH. The column density computed by integrating to the edge of the dust sublimation radius is $9.5 \times 10^{23}$ cm$^{-2}$ if the other fiducial parameters are fixed.}
\label{FiducialParameters}
\end{deluxetable}

\subsection{Initial gas configuration - clumpy models}
\label{ClumpySetup}

It has long been predicted on theoretical grounds that the dusty gas surrounding an accreting SMBH will not be smoothly distributed, but will instead form clumps \citep{Krolik1988}. This prediction has been supported by observations such as the variability of x-ray absorbing column densities in type 2 Seyferts \citep{Risaliti2002} as well as IR spectroscopy \citep{Mason2006, Honig2010a, Deo2011}. A vast literature exists concerning radiative transfer through clumpy torus models, with many prescriptions for generating clumpy density distributions from an underlying smooth density model and comparing the results to observations \citep{Nenkova2002, Elitzur2004, Honig2006, Schartmann2008, Stalevski2012, Heymann2012}. 

\begin{figure}
\includegraphics[width=0.5\textwidth]{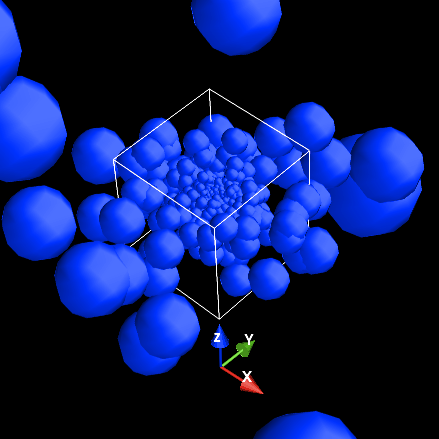}
\caption{A clumpy gas distribution corresponding to the fiducial parameters in Table \ref{FiducialParameters} and the clumping parameters for model 3 in Table \ref{ClumpingParameters}. Not pictured is the diffuse background gas. The white cube drawn at the center has a side length of 2 parsecs.}
\label{BuildingUpSnapshot}
\end{figure}

Our method for generating the clumpy gas distributions most closely resembles those of \citet{Honig2006} and \citet{Schartmann2008}. We use a three dimensional grid and spherical-polar $(r,\theta, \phi)$ coordinates, with logarithmic spacing in the radial coordinate and linear spacing in the angular coordinates. Our resolution is 128 radial zones, 96 $\theta$ zones for all $\theta$ ranging between $0$ and $\pi$, and 192 $\phi$ zones for all $\phi$ ranging from $0$ to $2 \pi$. The density of each clump in a given simulation is the same, and a preset number of clumps are placed on the grid. The clump positions are sampled from a probability distribution derived from a smooth density distribution as described in section \ref{DiskGeometry}. If two clumps overlap in position, their densities are added. Each clump's radius is set to a fixed number of grid zones in a given simulation, and the logarithmic radial spacing of the grid causes the size and optical depth of the clumps to grow with increasing distance from the SMBH. Overlaid on the clumps is a diffuse, smooth background gas distribution that is generated by multiplying the density distribution from section \ref{DiskGeometry} by 10$^{-2}$. An example is pictured in Figure \ref{BuildingUpSnapshot}.

For each clumpy gas distribution, the total mass in the computational domain was set equal to that of our fiducial smooth density model. The parameters varied between each clumpy gas distribution are the ratio of the clump diameter to the radial distance of the clump from the black hole ($d_{\rm cl}/r$), the gas density in the clump ($n_{\rm cl}$), and the number of clumps in the simulation volume $N_{\rm cl}$. The choices of $N_{\rm cl}$ and $d_{\rm cl}$ set the average number of clumps per line of sight, which we compute by averaging over all ($\theta$, $\phi$) sightlines with a weighting to account for the solid angle subtended by each sightline. 

We have chosen four combinations of clumping parameters to allow the average number of clumps per line of sight to take on values as low as 6 (in line with the results from \citet{Mor2009}) to as large as 105 in order to demonstrate a transition to the smooth density models. These parameter choices are listed in Table \ref{ClumpingParameters}.

\begin{deluxetable}{ccccc}
\tablecaption{Clumping parameters}
\tablehead{\colhead{Model \#} & \colhead{$d_{\rm cl}/r$} & \colhead{$n_{\rm cl}$ ( cm$^{-3}$ )} & \colhead{$N_{\rm cl}$} & \colhead{average number of} \\ \colhead{} & \colhead{} & \colhead{} & \colhead{} & \colhead{clumps per l.o.s.}}
\startdata
1 & .24 & $9.8 \times 10^4$ & 36864 & 105   \\
2 & .12 & $7.8 \times 10^5$ & 36864 & 25   \\
3 & .24 & $7.8 \times 10^5$ & 4608 & 13   \\
4 & .49 & $7.8 \times 10^5$ & 576 & 6.5  \\
\enddata
\tablecomments{See text for description of parameters}
\label{ClumpingParameters}
\end{deluxetable}

Ultimately we find that it is the volume filling fractions of the clumpy gas models that correlate most strongly with the integrated force exerted by the accretion radiation. If we let $f(r)$ denote the ratio of the volume occupied by at least one clump to the total volume within a sphere of radius $r$ centered on the black hole, then we find that it can be well approximated via broken power laws. For model 1, 
\begin{align}
f(r) &\approx 0.1 \times \left(\frac{r}{0.27 {\rm \; pc}}\right)^{-0.33} \qquad & {\rm for } \; 0.1 {\rm \; pc} < r < 2 {\rm \; pc} \; , \nonumber \\
  &\approx 0.05 \times \left(\frac{r}{2.3 {\rm \; pc}}\right)^{-1.5} \qquad & {\rm for } \; r > 2 {\rm \; pc} \; .
\end{align}
For models 2 through 4,
\begin{align}
f(r) &\approx 0.1 \times \left(\frac{r}{0.19 {\rm \; pc}}\right)^{-0.75} \qquad & {\rm for } \; 0.1 {\rm \; pc} < r < 1 {\rm \; pc} \;, \nonumber \\
  &\approx 0.03 \times \left(\frac{r}{1 {\rm \; pc}}\right)^{-1.5} \qquad & {\rm for } \; r > 1 {\rm \; pc} \; .
\end{align}

\begin{figure}

\begin{minipage}[b]{\linewidth}
\includegraphics[width=\textwidth]{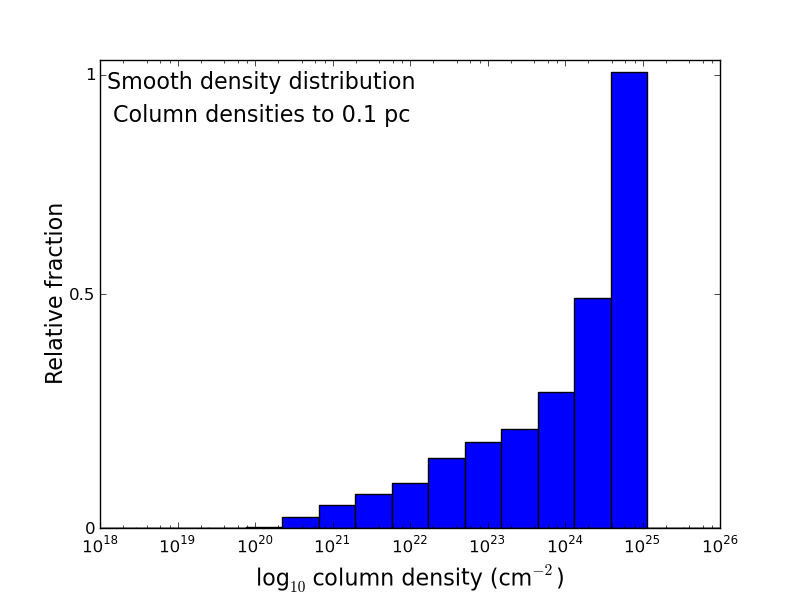}
\end{minipage}

\begin{minipage}[b]{\linewidth}
\includegraphics[width=\textwidth]{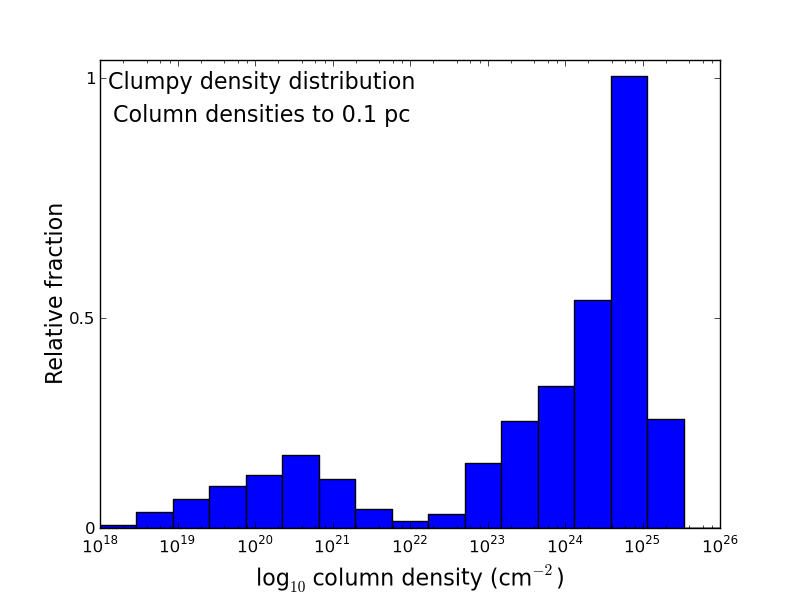}
\end{minipage}

\begin{minipage}[b]{\linewidth}
\includegraphics[width=\textwidth]{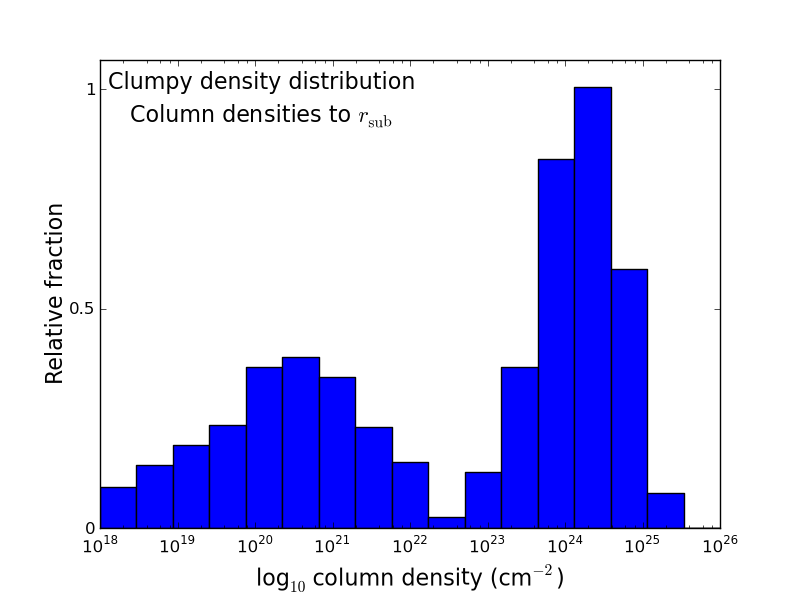}
\end{minipage}

\caption{Top: Column density histogram (integrated for all gas from 0.1 pc to large radii) for a smooth density model with the fiducial parameters listed in Table \ref{FiducialParameters}. Middle: Clumpy gas column density distribution with same clumping parameters as those used in Figure \ref{BuildingUpSnapshot}, also integrated for all gas from 0.1 pc to large radii. Bottom: All parameters are the same as the panel above, but this time the column density is integrated from the sublimation radius outward (i.e. these are the dusty gas columns).  }

\label{ColumnDensityHistograms}
\end{figure}

Figure \ref{ColumnDensityHistograms} shows the distribution of column density along randomly sampled sightlines for both our smooth and clumpy models. All column density values quoted in this study assume a mean mass per particle of 1.5 times the proton mass.

Making the gas clumpy leads to a larger number of sightlines with lower column densities compared to the smooth gas distribution, and spreads out the peak on the higher end of the column density distribution. Both of these effects are more in line with observational surveys of AGN \citep{Risaliti1999, Akylas2009, Malizia2009, LaMassa2009, Treister2009}. At the same time, our clumping prescription tends to make the column density distribution bimodal, with a division between sightlines that intersect no clumps versus those that intersect at least one clump. This bi-modality, which is not present in the observations, persists for all clumping parameters considered in this study, although it can be avoided if a larger fraction of the mass is allocated to the diffuse phase. 

\subsection{Monte Carlo Radiative Transfer}
The Monte Carlo technique partitions the luminosity of the accreting black hole into equal-energy photon packets that probabilistically interact with the surrounding gas. The packets were transported in three dimensions for all calculations in this study. We improve our statistics by mapping the energy and momentum deposited by the packets into a two-dimensional array of zones -- a photon that scatters at spherical coordinates ($r$,$\phi$,$\theta$) is mapped to position ($r$,$\theta^\prime$) where $\theta^\prime = \theta$ if $0 \le \theta \le \pi/2$ and $\theta^\prime = \pi - \theta$ if $\pi/2 < \theta \le \pi$.

In Monte Carlo radiative transfer, the specific intensity of the radiation $I(\boldsymbol{r},\hat{\boldsymbol{n}},\lambda)$ is constructed by counting the number of photon packets with wavelength $\lambda$ that enter into each grid zone at position $\boldsymbol{r}$ and with direction vector $\boldsymbol{n}$ in a given interval of time. Specifically, the radiation force per volume $f_\mathrm{rad}$ at a given position is defined as 
\begin{equation}
\boldsymbol{f}_\mathrm{rad} \equiv \frac{1}{c}\int_{\omega,\lambda} {\rho \, \kappa_\lambda \, I \, \hat{\boldsymbol{n}} \, d\omega \, d\lambda} \; .
\label{RadiativeAcceleration}
\end{equation}

To compute $f_\mathrm{rad}$ in a given zone of our computational domain with volume $\Delta V$ over a time interval $\Delta t$, we perform a sum a sum over all photon packets entering the zone. Each photon packet carries with it an energy $E_p$, a direction of travel $\hat{\boldsymbol{n}}_p$, and a wavelength $\lambda_p$. Associated with that wavelength is an opacity $\kappa(\lambda_p)$, measured per gram of gas, and which depends on whether dust is present at location $\boldsymbol{r}$. The packet traverses a path of length $\Delta r$ within a zone at position $\boldsymbol{r}$. The force is then
\begin{equation}
\boldsymbol{f}_\mathrm{rad} = \left(\frac{1}{\Delta \, V \Delta \, t}\right) \, \rho(\boldsymbol{r}) \, \sum_p \frac{E_p}{c} \, \kappa(\lambda_p, \boldsymbol{r})\,  \Delta r \,  \hat{\boldsymbol{n}}_p \;.
\end{equation}
The radiative acceleration $a_\mathrm{rad}$ is simply defined as $f_\mathrm{rad}/\rho$.

Our calculations apply the stationarity approximation, in which we solve the steady-state radiative transfer problem for a fixed gas density distribution. This approximation is justified if the radiative heating time scale and the radiative diffusion time scale are much shorter than the dynamical time scale. 

For a sound speed of 200 km s$^{-1}$ and a characteristic length scale of 10 pc, the dynamical time is approximately $10^{12}$ seconds. Meanwhile, the photon diffusion time through the disk never exceeds $10^{11}$ seconds, and for many disk parameters the diffusion time is substantially shorter than that. The radiative heating time, estimated by dividing the thermal energy of the gas by the rate of radiative energy deposition, is
\begin{align}
&t_\mathrm{heat} \approx \left(\frac {\rho \, k_B \, T_\mathrm{gas}}{\mu m_p}\right) \left( \frac{1}{\rho \, \kappa \, c \, a \, T_{\mathrm{rad}}^4  } \right) \nonumber \\
&= 2.4 \times 10^5 \mathrm{\; s} \left(\frac{T_\mathrm{gas}}{100 \mathrm{\; K}} \right) \left(\frac{T_\mathrm{rad}}{100 \mathrm{\; K}} \right)^{-4} \left(\frac{\kappa}{10 \mathrm{\; cm}^2 / \mathrm{g}} \right)^{-1} \;,
\end{align}
which is also much shorter than the dynamical time.

In this case, the condition of radiative equilibrium allows us to compute the dust temperatures by balancing radiative heating and cooling,
\begin{eqnarray}
&4 \pi \int_\lambda { \rho \,  \kappa_\mathrm{abs} (\lambda) \, B_\lambda (T_\mathrm{dust}) \, d\lambda} \nonumber \\ &= \int_{\omega,\lambda}{\rho \,\kappa_\mathrm{abs} (\lambda) I_\lambda \, d\omega \, d\lambda} \;.
\end{eqnarray}

For most calculations, the photons are emitted isotropically at the edge of the 0.1 pc sphere surrounding the origin. The effect of anisotropic emission is treated in section \ref{AnisotropicResults}. We follow the photon propagation for time intervals of $5 \times 10^9$ seconds, at which point we update the temperature of the dust in each grid zone. We treat dust as present everywhere where the dust temperature is below 1400 Kelvin. The dust temperatures are updated until convergence is obtained at the one percent level, which typically takes fewer than 40 iterations if the initial dust temperature is set to 100 Kelvin in every zone. 

Finally, for estimating the dynamics of the gas based on the radiation pressure on the dust, we assume perfect hydrodynamical coupling between the dust and the gas, as justified in \citet{Murray2005}. 

\subsection{Intrinsic AGN spectrum}
We use the ``intrinsic'' (unreddened) AGN spectral energy distribution described in \citet{Marconi2004}. The majority of the spectral energy is found in the optical and near-UV and originates from the accretion disk, which resembles a $10^5$ Kelvin black body emitter. The spectrum also contains a sizable x-ray component. Intentionally absent from this spectrum is any infrared component, which we will calculate self-consistently based on the reprocessing of the radiation by dust.

\label{IntrinsicSpectrum}
\subsection{Dust and electron interactions}
\label{DustAndElectronInteractions}

We use tabulated dust opacities and albedos based on \citet{Draine2003-1} for wavelengths greater than 10 Angstroms, and \citet{Draine2003-2} for shorter wavelengths, all corresponding to visual extinction ratio $R_V= 3.1$ and assuming a fixed dust-to-gas mass ratio of 1/125. These values were interpolated between 48 reference wavelengths. In practice, the difference between scattering and absorption is that for an absorption interaction, the wavelength of the re-emitted photon packet will be sampled from a probability distribution that depends on the dust's temperature, whereas the wavelength will remain unchanged for a scattering interaction. For wavelengths less than 100 Angstroms, we ignore scattering by dust since it will be almost entirely in the forward direction and hence will not lead to a net transfer of momentum, although we still allow for absorption by dust. 

Electron scattering is only relevant for photons with wavelengths less than $\sim 10$ Angstroms, when the dust absorption cross section drops below that of the Thompson cross section, and when the photons are energetic enough to scatter equally well off of both bound and free electrons. We account for anisotropic, inelastic electron scattering in accordance with the Klein-Nishina formula. 

\section{Results}
\label{Results}

\subsection{Dust temperature and radiative acceleration dependence on smooth gas geometry}
\label{TemperatureAndAccelerationResults}

\begin{figure}

\begin{minipage}[b]{\linewidth}
\includegraphics[width=\textwidth]{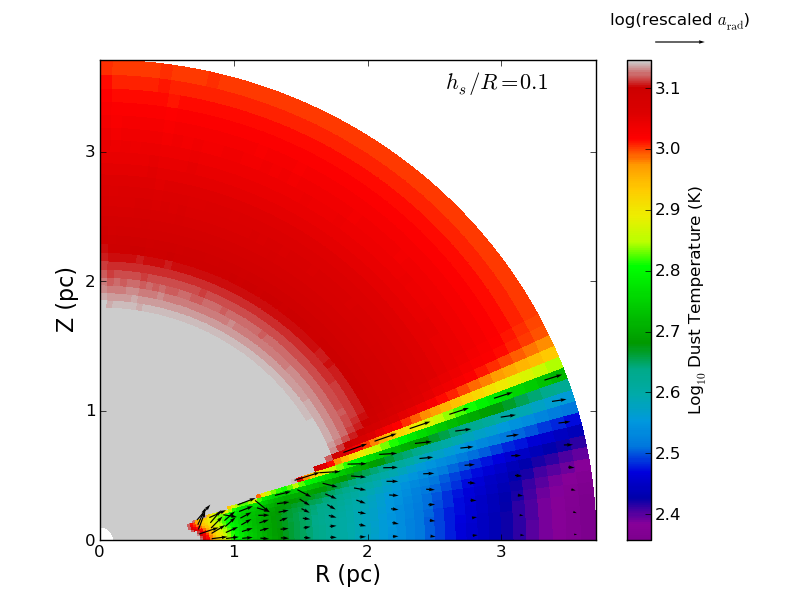}
\end{minipage}
\begin{minipage}[b]{\linewidth}
\includegraphics[width=\textwidth]{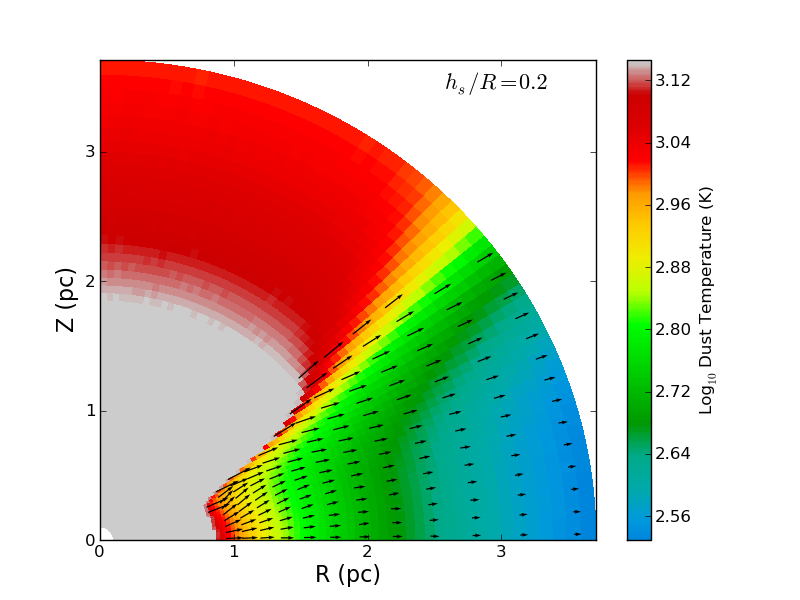}
\end{minipage}
\begin{minipage}[b]{\linewidth}
\includegraphics[width=\textwidth]{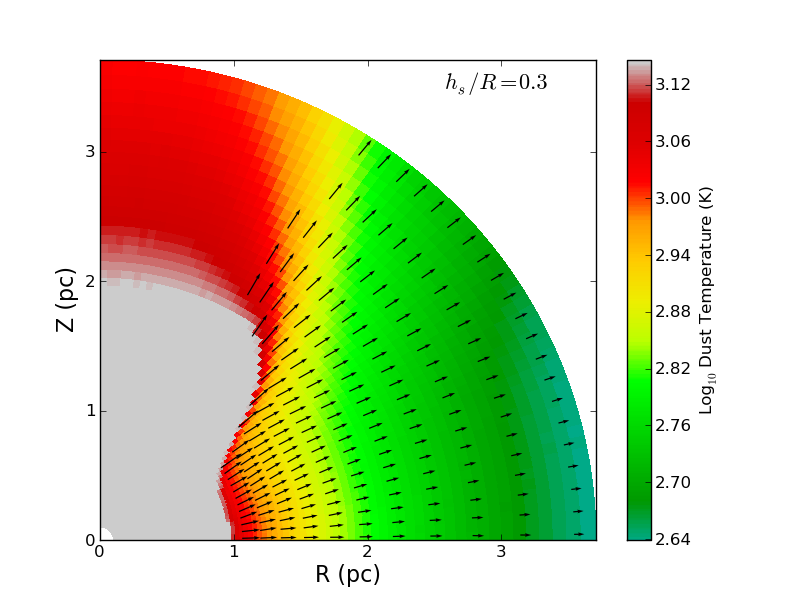}
\end{minipage}
\caption{Arrows indicating the direction and strength of the radiative acceleration are plotted over slices of dust temperature. All parameters correspond to the fiducial values in Table \ref{FiducialParameters}, except for opening angles which vary as indicated (while conserving mass in the calculation domain). Regions in gray indicate where dust is sublimated (dust temperature that exceeds 1400 K). Acceleration arrows are present in zones where the dust is not sublimated and the gas density exceeds $10^{-21}$ g / cm$^3$. The arrow lengths are proportional to log$_{10}$($10^6 \times a_\mathrm{net}$) where $a_\mathrm{net}$ is in cgs units.}
\label{DustTempSlices}
\end{figure}

Figure \ref{DustTempSlices} shows slices of the equilibrium dust temperature and the radiative acceleration vector field for disks of varying opening angles and with a smooth gas distribution. The color scheme is set so that all temperatures above the dust sublimation temperature appear as solid gray. Arrows representing the acceleration are plotted in zones where the dust is not sublimated and where the gas density exceeds $10^{-21}$ g cm$^{-3}$.

We find that the dust sublimation region has an aspherical, hour-glass shape. Sublimation extends to larger radii in the polar regions where the dusty gas is optically thin in the infrared. There, the dust absorbs ultraviolet radiation but emits in the infra\
red, forcing it to reach a higher temperature to maintain radiative equilibrium.

\begin{figure}
\includegraphics[width=0.5\textwidth]{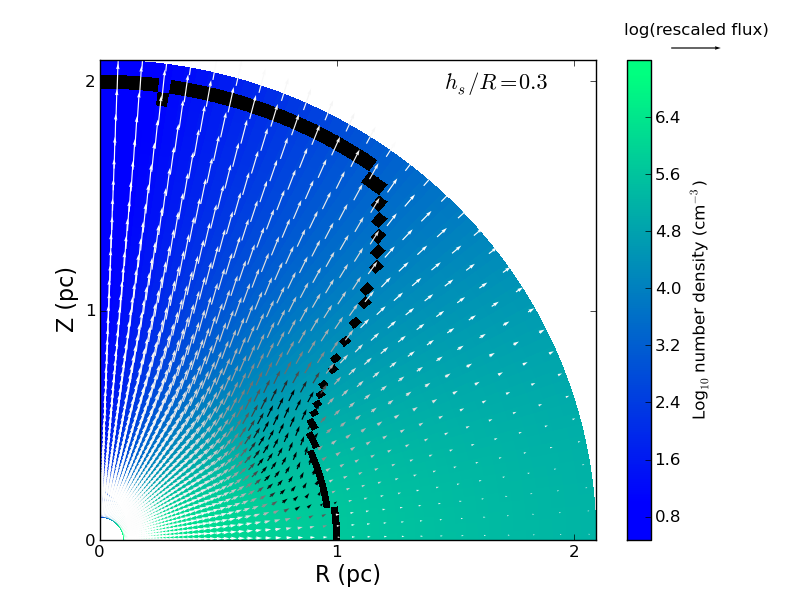}
\caption{Arrows indicating the radiation flux are plotted over gas density. All parameters in this calculation correspond to the fiducial values listed in Table \ref{FiducialParameters}. Arrows with significant deviation from the radial direction are colored black, while the boundary of the dust sublimation region is marked with black cells. The arrow lengths are proportional to log$_{10}$($10^{-14} \times$ net flux (cgs)). Through a process of absorption of UV light and re-emission in the IR at the inner wall of the dusty gas, flux is channeled toward the poles in the outermost part of the dust sublimation region. The radiation travels radially in the dusty portion of the gas.}
\label{FluxSlice}
\end{figure}

Interestingly, nearly all the angular redistribution of the radiation occurs near the surface of the dust sublimation region. Light from the central source initially travels isotropically to the inner edge of the dusty gas, and a large fraction of the photons are absorbed at the dust interface. When photons are re-emitted in the infrared, many are sent back into the sublimation region. It is through this re-emission that the net flux becomes anisotropic at small radii. When infrared photons succeed in penetrating deep into the dusty gas, they generate a nearly radial radiative flux, as they would in a spherically symmetric problem (see Figure \ref{FluxSlice}).

\begin{figure*}
\centering
\includegraphics[width=0.75\textwidth]{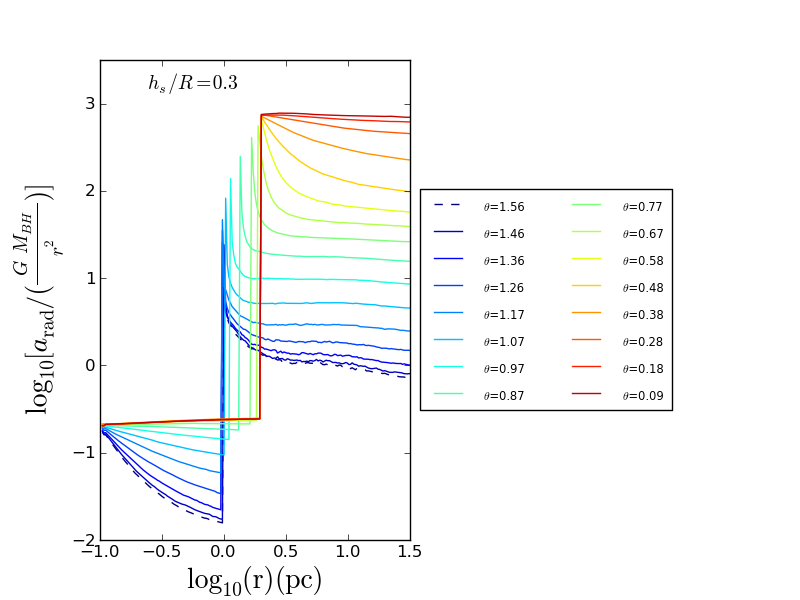}
\caption{Radiative acceleration in radius and solid angle. All parameters in this calculation correspond to the fiducial values listed in Table \ref{FiducialParameters}. The acceleration is normalized at each radius by the gravitational acceleration $G  M_\mathrm{BH} / r^2$, and the logarithm of that ratio is plotted. Thus, points with y-values above zero correspond to locations where the radiative acceleration exceeds gravity. The abrupt jump in acceleration occurs at the boundary of the dust sublimation region, where dust begins to contribute to the radiative opacity. As the radius increases beyond this boundary, the mean wavelength of the radiation transitions from the UV to the IR, rapidly lowering the radiative opacity in the process and reducing the radiative acceleration until eventually obeying an inverse square law dependence on radius.}
\label{RadAccelInRadiusNormalized}
\end{figure*}

Figure \ref{RadAccelInRadiusNormalized} displays how the radiative acceleration varies with radius and polar angle for the fiducial simulation. The behavior of the acceleration is quite different inside and outside the dust sublimation region -- the presence of dust raises the opacity of the gas and therefore raises the radiative acceleration (as in equation \ref{RadiativeAcceleration}). In a given solid angle the acceleration is highest just beyond the edge of the dust sublimation region, where ultraviolet and optical photons can push on optically thick, dusty gas. The acceleration rapidly drops as the radiation penetrates farther into the dusty gas and ultraviolet/optical light is converted into infrared, to which the dust is less opaque. For all solid angles, the acceleration settles to a constant ratio above gravity at sufficiently large radius, indicating that the acceleration eventually becomes proportional to $1 / r^2$, further evidence that the infrared radiation diffuses primarily in the radial direction. In addition to the radial dependence of the acceleration, there is an angular dependence that arises from the diversion of flux from the mid-plane to the polar regions of the disk at the surface of the sublimation region. 

\begin{figure}
\begin{minipage}[b]{\linewidth}
\includegraphics[width=\textwidth]{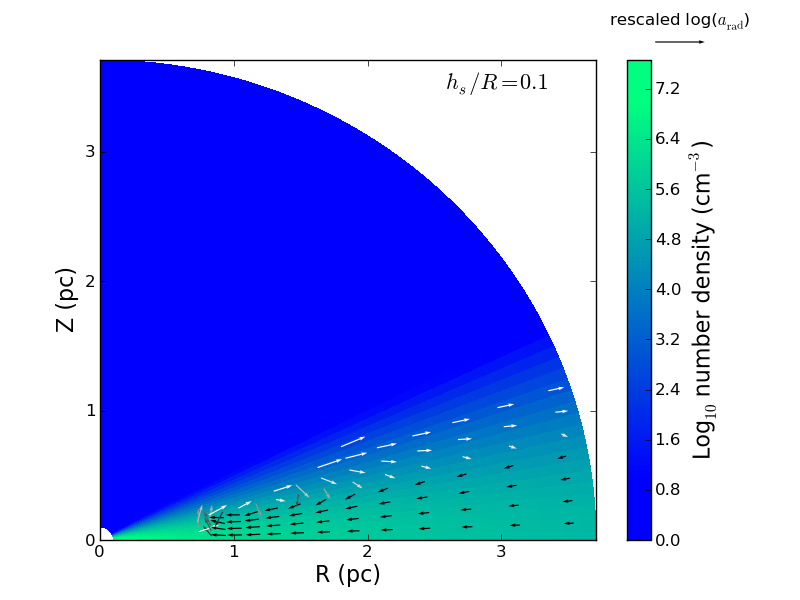}
\label{CombinedAccelSlices:hoverRp1}
\end{minipage}
\begin{minipage}[b]{\linewidth}
\includegraphics[width=\textwidth]{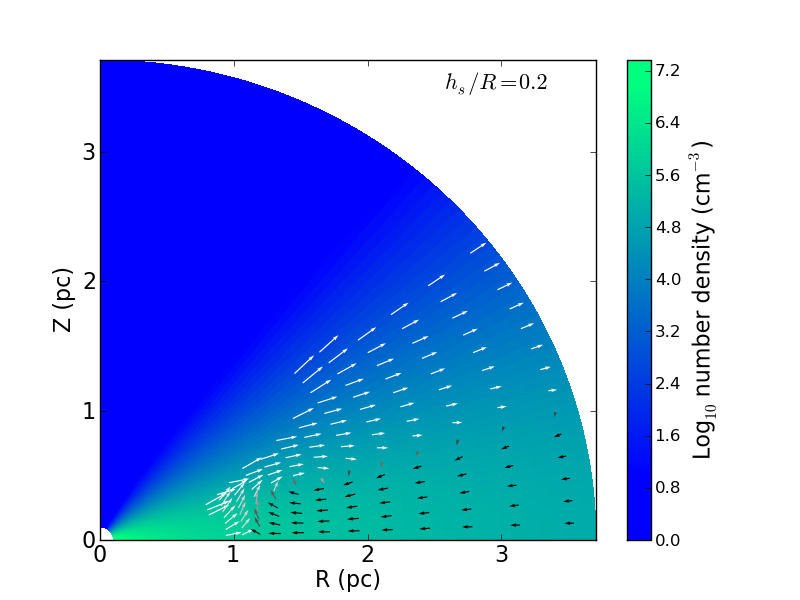}
\label{CombinedAccelSlices:hoverRp2}
\end{minipage}
\begin{minipage}[b]{\linewidth}
\includegraphics[width=\textwidth]{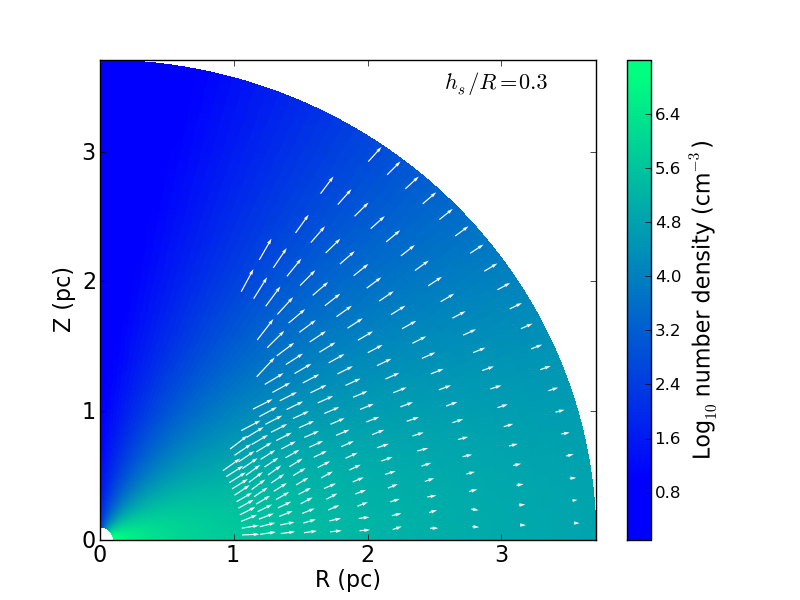}
\label{CombinedAccelSlices:hoverRp3}
\end{minipage}
\caption{Arrows representing net acceleration (radiation + gravity) as a function of position. All parameters for this calculation correspond to the fiducial values listed in Table \ref{FiducialParameters}, except for opening angles which vary as indicated (while conserving mass in the calculation domain). Inward-directed arrows are colored black, and the arrow lengths are proportional to log$_{10}$($10^6 \times a_\mathrm{net}$) where $a_\mathrm{net}$ is in cgs units. For small opening angles, the gravitational acceleration exceeds that of the radiation in the equatorial region up to a critical angle above the mid-plane, and beyond this angle radiation dominates. As the opening angle increases, photons deposit more momentum in the dusty gas, and for sufficiently large polar angle the radiation force can exceed gravity in all directions.}
\label{CombinedAccelSlices}
\end{figure}

Slices of the net acceleration with gravitational acceleration included are shown in Figure \ref{CombinedAccelSlices}. In all cases the acceleration is primarily radial in direction, either outward or inward. Note that for opening angles $h_s/R < 0.3$ there is a critical polar angle below which radiation dominates over gravity and above which gravity dominates. In these cases inflow may persist in the equatorial region while gas is blown out at angles directed farther away from the mid-plane, potentially leading to a steady state outflow. However, the radiative acceleration dominates over gravity everywhere when $h_s/R \ge 0.3$ for this AGN luminosity and column density.

\begin{figure}
\includegraphics[width=0.5\textwidth]{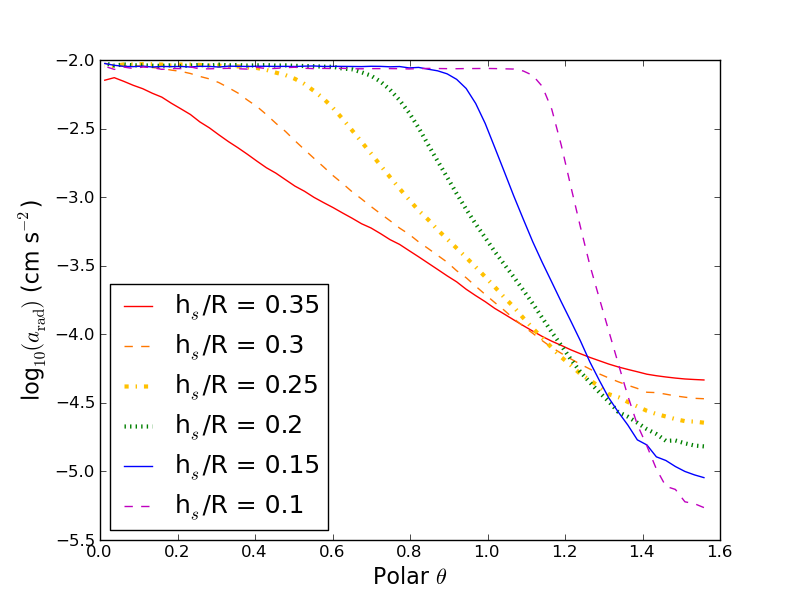}
\caption{Radiative acceleration as a function of polar angle (gravity not included). All parameters for this calculation correspond to the fiducial values listed in Table \ref{FiducialParameters}, except for opening angles which vary as indicated (while conserving mass in the calculation domain). The acceleration is lowest in the equatorial region, even though the force from radiation pressure is highest there, because the force does not rise as quickly as the mass as the polar angle increases.}
\label{CombinedAccelInAngle}
\end{figure}

For another perspective, in Figure \ref{CombinedAccelInAngle} we plot the integrated radiative acceleration for columns of gas as a function of polar angle (without gravitational acceleration included). We assume there are no forces in the tangential directions (i.e., each column is accelerated independently), and that the radiation force is shared along the whole column as the inner gas pushes on outer gas. To compute this net acceleration, we first compute the integrated net force in each solid angle (including the effects of both gravity and radiation),

\begin{equation}
\frac{ d F_\mathrm{net}}{ d \omega }(\theta) \equiv \int_{r_\mathrm{sub}} ^{r_\mathrm{out}}{\left( f_\mathrm{rad} - \frac{G \, M_\mathrm{BH} \, \rho}{r^2}\right)r^2 \, dr } \; ,
\end{equation}

along with the mass in that solid angle,
\begin{equation}
\frac{d M_\mathrm{gas}}{d \omega}(\theta)  \equiv \int_{r_\mathrm{sub}} ^{r_\mathrm{out}} {\rho \, r^2  \, dr } \;,
\label{NetAccelerationIntegral}
\end{equation}
where $r_\mathrm{sub}$ denotes the edge of the dust sublimation region for each value of $\theta$. Then, the net integrated acceleration is simply
\begin{equation}
a_\mathrm{net}(\theta) \equiv \left [ \frac{d F_\mathrm{net}}{d \omega} \right ] \big/ \left[\frac{d M_\mathrm{gas}}{d \omega}\right] \;.
\end{equation}

The value of $a_\mathrm{net}$ depends on the choice of $r_\mathrm{out}$. However, we will show in section \ref{MdotScaling} that the dependence of the rate of mass outflow on $r_\mathrm{out}$ is very small.

From Figure \ref{CombinedAccelInAngle} we see that as the opening angle of the parsec-scale disk becomes smaller, the radiative acceleration becomes more sharply divided between the optically thin and optically thick portions of the disk. This primarily reflects the sharper density gradients present for smaller opening angles. As we will show in section \ref{EffectiveTauOpeningAngles}, even though the radiative force is greater in the optically thick portion of the disk, the force does not rise as quickly as the mass. This causes the acceleration to decrease toward the mid-plane.

\subsection{Enhancement of radiation force above $L/c$ and the dependence on smooth gas geometry}
\label{EffectiveTauOpeningAngles}

The ability of radiation to clear away ambient gas is enhanced by the fact that diffusing photons deposit their momentum multiple times as they random walk outwards. We can quantify this effect in each solid angle by dividing the integrated force on the gas column in that solid angle by the radiative momentum per time per solid angle leaving the inner source. We call the resulting quantity $\tau_\mathrm{eff}(\theta)$, which is computed as
\begin{equation}
\tau_\mathrm{eff}(\theta) =  \left[\frac{ d F_\mathrm{rad}}{ d \omega }\right] \big/ \left[\left(\frac{1}{4 \pi} \right) \frac{L_\mathrm{BH}}{c} \right] \; ,
\label{TauEffDefinition}
\end{equation}
and we extend the lower limit of the integral defining $d F_\mathrm{rad}/ d \omega$ from $r_\mathrm{sub}$ to $0$ when computing this quantity. We may also compute an average value of this quantity averaged over all lines of sight,
\begin{equation}
\overline{\tau}_\mathrm{eff} \equiv  \frac{1}{4 \pi} \int_{\omega} \tau_\mathrm{eff} \, d \omega  =  \int_{0} ^{\pi/2} \tau_\mathrm{eff} \, \sin \theta \, d \theta \;.
\end{equation}

\begin{figure}
\includegraphics[width=0.5\textwidth]{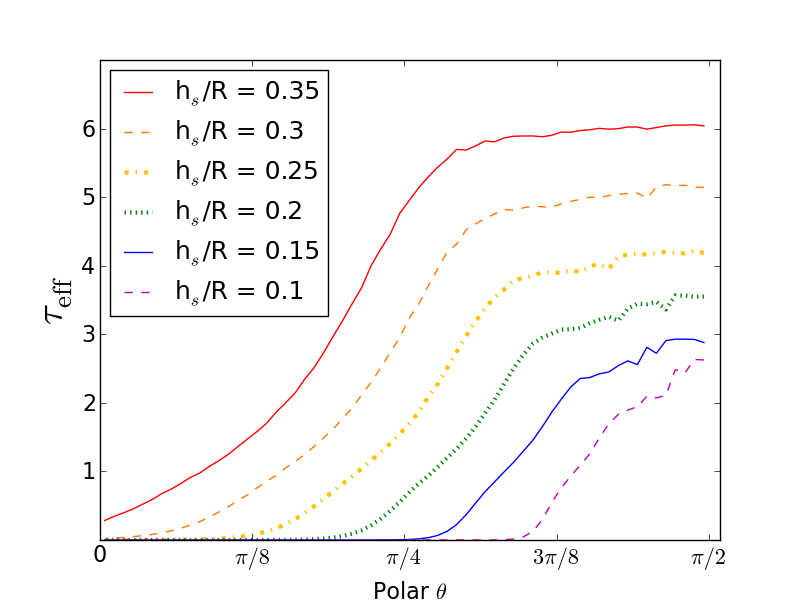}
\caption{$\tau_{\mathrm{eff}}(\theta)$ for various opening angles (see equation \ref{TauEffDefinition}). Parameters correspond to the fiducial values in Table \ref{FiducialParameters}, except for opening angles which vary as indicated (while conserving the total mass in the calculation domain). $\tau_{\mathrm{eff}}(\theta)$ is a measure of the radiative force on the gas column at a given polar angle, and it reaches its highest values in the equatorial region.}
\label{TauOpeningAngles}
\end{figure}

Figure \ref{TauOpeningAngles} summarizes our results for $\tau_\mathrm{eff}$ for various disk opening angles while holding the other parameters at their fiducial values as listed in Table \ref{FiducialParameters}. Increasing the opening angle boosts $\tau_\mathrm{eff}(\theta)$ for all $\theta$, up to a maximum value of approximately 5-6 for these parameters. In the polar region, this effect can be understood simply in terms of the presence of more mass in that region when the opening angle is larger. Meanwhile, even though there is less mass present in the equatorial region as the opening angle increases, the radiative flux in that region increases such that $\tau_\mathrm{eff}(\theta)$ is able to increase there as well. 

If we calculate the radiation force on spherically distributed gas with the same $\overline{N}_\mathrm{H}$, we find that $\tau_\mathrm{eff}$ = 13. Thus, even though the effective radiation force exceeds $L/c$ in Figure \ref{TauOpeningAngles}, the enhancement is not as large for a realistic disk geometry as it is in the spherically symmetric case. For the largest opening angle considered in this study ($h_s/R = 0.35$), $\overline{\tau}_\mathrm{eff}$ is smaller than the spherically symmetric value by a factor of $\sim 2$.

\subsection{Results for Clumpy Gas}
\label{ClumpyResults}

\begin{figure}
\includegraphics[width=0.5\textwidth]{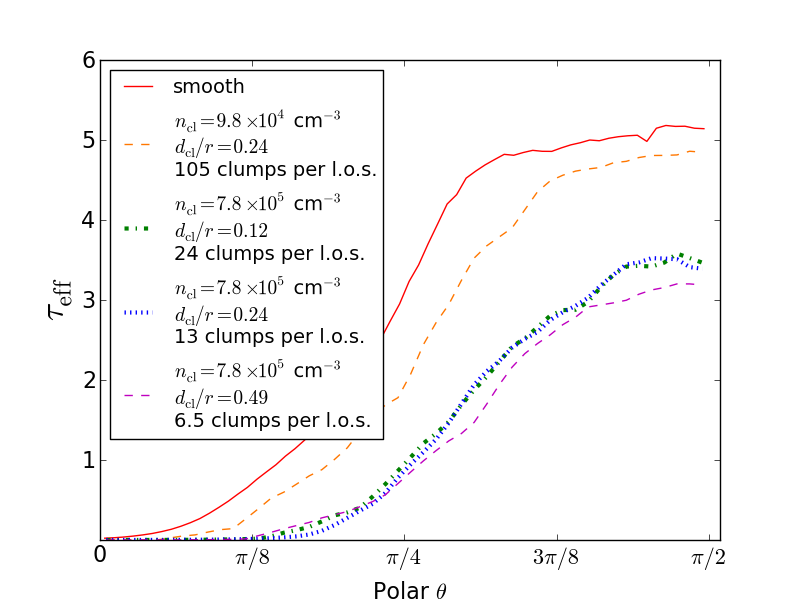}
\caption{$\tau_\mathrm{eff}(\theta)$ for various clump densities $n_\mathrm{cl}$ and clump sizes. Clump sizes are specified by the ratio of the clump diameter $d_\mathrm{cl}$ to clump radial position $r$. In all simulations the number of clumps is varied such that the total mass of the gas in the simulations domain is held constant. The diffuse background makes up 1\% of the mass in all simulations. The number of clumps per line of sight, listed in the legend, is computed by averaging over all ($\theta$, $\phi$) sightlines with a weighting to acount for the solid angle subtended by each sightline.  }
\label{TauClumpySphereComparison}
\end{figure}

Figure \ref{TauClumpySphereComparison} shows how $\tau_\mathrm{eff} (\theta)$ varies with the clumpiness of the gas. The shape of the momentum deposition as a function of $\theta$ appears generally the same for the clumpy and smooth cases. This suggests that the smooth density distributions employed throughout most of this study provide accurate approximations to the behavior of more realistic clumpy density distributions, although they should systematically overestimate the radiation force on the dusty gas by a factor of $\sim 2$ in the most extreme cases of clumping (fewest clumps per line of sight) considered here.

\begin{figure}
\includegraphics[width=0.5\textwidth]{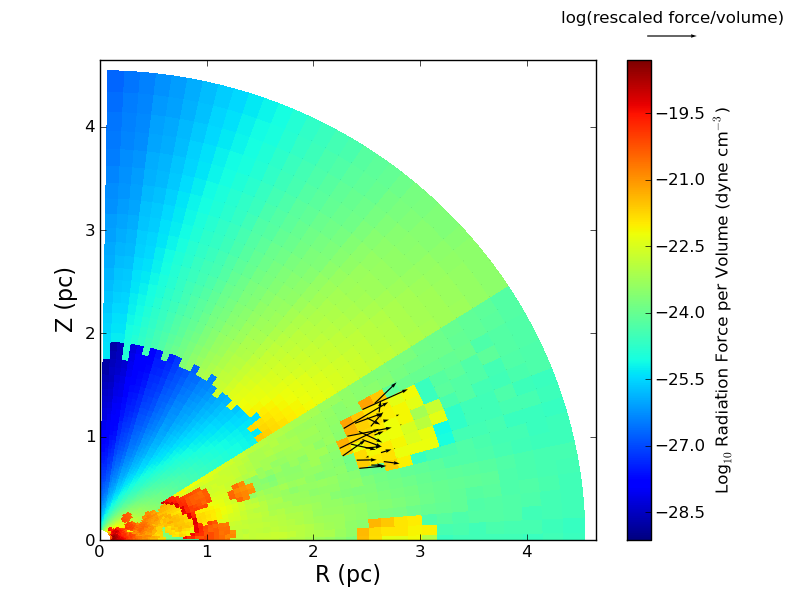}
\caption{A slice of the magnitude of the radiation force for a clumpy gas distribution. A jump in the magnitude of the force is evident at the dust sublimation boundary. Arrows indicating the direction and magnitude of the radiation force are overlaid on one clump. The arrow lengths are proportional to log$_{10}(1.5 \times 10^{22}) \times f_\mathrm{rad}$ (cgs).}
\label{ForceOnClumps}
\end{figure}

Figure \ref{ForceOnClumps} illustrates how the radiation force acts on portions of individual clumps. Note how the force remains radially directed even in the presence of clumps, and how clumps shadow gas behind them.

\subsection{Results for Anisotropic AGN Emission}
\label{AnisotropicResults}

If the black hole accretion disk is aligned with the mid-plane of the gas present at the scale of our calculation, one might expect that there would be more flux emitted in the polar directions than in the mid-plane direction. According to one prescription \citep{Netzer1987}, the emitted flux should obey
\begin{equation}
F_\mathrm{emitted} \propto \cos \theta \, ( 1 + 2 \cos \theta) \;,
\label{AnisotropicEmission}
\end{equation}
where the first factor accounts for projected surface area and the second factor accounts of limb-darkening in an optically thick
atmosphere. There is reason to doubt the validity of this model when relativistic effects are taken into account which tend to redirect radiation back toward the mid-plane \citep{Sun1989}. Moreover, it remains unclear whether the black hole accretion disk is aligned with the torus. We nevertheless choose to explore the scenario described by equation \ref{AnisotropicEmission} in order to test the sensitivity of our results to variations in the emission pattern of the accretion disk.

Unlike the case of isotropic emission, we find that for anisotropic emission $\tau_\mathrm{eff}(\theta)$ peaks at an intermediate angle $< \pi/2$. The peak arises because at small polar angles there is hardly any gas present to provide optical depth, whereas hardly any light penetrates into the dusty gas at large polar angles. For a calculation with our fiducial parameters, the ratio of $\overline{\tau}_\mathrm{eff}$ in the case of anisotropic emission versus $\overline{\tau}_\mathrm{eff}$ for the case of isotropic emission is 0.72, indicating that photons tend to escape from the disk with fewer interactions when they are emitted in an anisotropic manner. This ratio will be even smaller for smaller disk opening angles.

If, instead of being diverted away form the torus plane as prescribed by equation \ref{AnisotropicEmission}, the radiation is beamed toward the plane, $\tau_{\mathrm{eff}}$ would \emph{increase} compared to the fiducial simulation.

From this point on, we will only consider models with a smooth density distribution and isotropic central emission. Nevertheless, it is important to keep in mind that the integrated force and $\dot{M}$ are likely to be modified due to the effects of gas clumping and anisotropic emission of radiation.

\subsection{Estimating the mass outflow rate}
\label{MdotScaling}
We cannot determine precisely the dynamics of the gas without coupling the radiative transfer calculation to a hydrodynamic solver in a time-dependent calculation. However, we may apply an Eddington-type argument to approximate whether gas will be blown away in a given solid angle and to estimate the mass-loss rate. This argument considers the gravitational and radiation forces but ignores centrifugal acceleration of the gas, viscous or gravitational torques, and shocks. The neglect of centrifugal support will not significantly alter the results in the cases when the radiation force on a column of gas is much less than or much greater than the corresponding force of gravity, but it will contribute to an under-estimation of the mass outflow rate in the intermediate range. 

Let $\overline{t} (\theta)$ denote the time taken to accelerate the gas in a given column with mass $d M_\mathrm{tot} $ to a distance $r_\mathrm{out}$ at constant acceleration $a_\mathrm{net} (\theta)$. Then 
\begin{equation}
\overline{t}(\theta) \approx \sqrt{\frac{2 \, r_\mathrm{out}}{a_\mathrm{net}}} = \sqrt{ 2 \, r_\mathrm{out} \frac{\frac{d M_\mathrm{gas}}   {d \omega }}{\frac{ d F_\mathrm{net}} { d \omega}}} \;.
\label{OutflowTime}
\end{equation} 

We define a differential mass outflow rate per solid angle $d \dot{M}/d \omega$,
\begin{equation}
\frac{d \dot{M}}{d \omega} (\theta) \equiv \frac{\frac{d M_\mathrm{gas}} {  d \omega}}{{\overline{t}(\theta)}}
= \sqrt{ \frac{ \left( \frac{d M_\mathrm{gas}} {d\omega} \right) \left( \frac{d F_\mathrm{net} }{ d\omega }\right) }{2 \, r_\mathrm{out}} } \;.
\label{MdotDefinition}
\end{equation}

We can also define a mass outflow rate integrated over the entire volume (all of $\theta$),
\begin{equation}
\dot{M} \equiv  \int_{\omega} { \frac{d \dot{M} }{ d \omega \,} \, d \omega } = 2 \, (2\pi) \int_0 ^{\pi/2} { \frac{d \dot{M} }{ d \omega \,} \sin \theta \, d \theta } \; ,
\label{IntegratedMdotDefinition}
\end{equation}
where we have taken advantage of the assumed symmetry for $\theta \rightarrow \pi - \theta$. Whenever $d\dot{M}/d\omega$ is less than zero for a particular value of $\theta$, we do not add it to the total reported value for the total volume-integrated $\dot{M}$, since we are only interested in the gas that gets blown away.

Our gas density prescription (section \ref{DiskGeometry}) indicates that for any given polar angle, the density goes as $r^{-\gamma}$. This allows us to compute $d M_\mathrm{gas}/d \omega$ in terms of $r_\mathrm{out}$ and the sublimation radius $r_\mathrm{sub}(\theta)$,  

\begin{eqnarray}
&\frac{d M_\mathrm{gas}}{d \omega} = \int_{r_\mathrm{sub}} ^{r_\mathrm{out}} { \rho(r_\mathrm{sub}) \, \left(\frac{r}{r_\mathrm{sub}}  \right )^{-\gamma} \, r^2 \, dr } \nonumber \\ 
&=  \frac{1}{3 - \gamma} \, \rho (r_\mathrm{sub}) \, r_\mathrm{out}^3  \, \left( \frac{ r_\mathrm{sub}}{r_\mathrm{out}} \right)^\gamma \left[1 - \left(\frac{r_\mathrm{sub}}{r_\mathrm{out}} \right)^{3-\gamma} \right] \; .
\label{MassInSolidAngle}
\end{eqnarray}

In section \ref{MdotOpeningAngleResults} we will present values for $d \dot{M}/d\omega$ calculated using equations \ref{MdotDefinition}, \ref{MassInSolidAngle}, and the values of $d F_\mathrm{net}/d\omega$ calculated using the Monte Carlo. For the rest of this section, we present a simple scaling argument to demonstrate that our estimates of the mass outflow rate depend only weakly on our choice of the outer radius $r_\mathrm{out}$ (which is somewhat arbitrary). 

Our results from section \ref{TemperatureAndAccelerationResults} indicate that we can think of the radiative acceleration as being divided into two parts: a spike in acceleration at the sublimation radius that arises from the absorption of ultraviolet and optical photons, and acceleration due to absorption of infrared photons that goes as $r^{-2}$ at large radii. Only the second type of acceleration is sensitive to our choice of $r_\mathrm{out}$. We may approximate the infrared radiation force as 
\begin{eqnarray}
& {\frac{ d F_\mathrm{net}}{ d \omega }}_{\mathrm{IR}}\approx \nonumber \\ & \int_{r_\mathrm{sub}} ^{r_\mathrm{out}} {  \rho(r_\mathrm{sub}) \left(\frac{r}{r_\mathrm{sub}}  \right )^{-\gamma}  \left[ \frac{a_\mathrm{rad}(r_\mathrm{sub}) r_\mathrm{sub}^2   - G \, M_\mathrm{BH}}{r^2} \right] r^2dr   } \nonumber \\  & = \frac{1}{\gamma - 1} \, \rho(r_\mathrm{sub})\left[ a_\mathrm{rad}(r_\mathrm{sub})  - \frac{G \, M_\mathrm{BH}}{r_\mathrm{sub}^2} \right] r_\mathrm{sub}^3 \nonumber \\ & \times \left[ 1  - \left( \frac{r_\mathrm{sub}}{r_\mathrm{out}}\right)^{\gamma - 1}  \right]   \;,
\end{eqnarray}
where $a_\mathrm{rad}(r_\mathrm{sub})$ refers to the value of the radiative acceleration at the sublimation radius that provides the correct normalization for the inverse-square law acceleration at large radii. 

Using equation \ref{MdotDefinition}, dropping factors of order unity, and assuming $r_\mathrm{out} \gg r_\mathrm{sub}$, we finally arrive at
\begin{eqnarray}
\label{FullMdotResult}
& {\frac{d \dot{M}}{d \omega}}_{IR} \approx  \nonumber \\ & \rho(r_\mathrm{sub}) \left[ a_\mathrm{rad}(r_\mathrm{sub})  - \frac{G \, M_\mathrm{BH}}{r_\mathrm{sub}^2} \right]^{1/2} r_\mathrm{sub}^{5/2}  \left(\frac{r_\mathrm{out}}{r_\mathrm{sub}}\right)^{1 - \frac{1}{2}\gamma} \;.
\end{eqnarray}

From the simulations performed in \citet{Hopkins2012-1}, $\gamma$ tends to fall between 1.5 and 2, and as already noted we have fixed $\gamma$ at 1.5 for all numerical calculations in this study. We expect that the density profile will ultimately truncate at about 1 kpc. So, the ratio of the $d \dot{M}/d \omega$ due to absorption of infrared photons that we would calculate using $r_\mathrm{out}$ of 1 kpc versus $r_\mathrm{out}$ of 32.4 pc would be, for $\gamma = 1.5$, only 2.4. For $\gamma =2$, $d \dot{M}/d \omega$ would be invariant with respect to choice of $r_\mathrm{out}$, and for $\gamma = 2.5$ the ratio would be 0.42. The fact that a significant portion of the radiative acceleration in the Monte Carlo calculations comes from the spike near the dust sublimation radius further reduces the sensitivity of our results to our choice of $r_\mathrm{out}$. 

A final quantity that will be useful to us is the velocity of the gas in a solid angle $v_\mathrm{out}(\theta)$, 
\begin{equation}
v_\mathrm{out}(\theta) \equiv \sqrt{2\, a_\mathrm{net} \, r_\mathrm{out}} \;.
\label{VoutDefinition}
\end{equation}

Once again focusing on the infrared acceleration at large radii and making the same approximations as we did for estimating $d \dot{M}/d \omega$, we find
\begin{eqnarray}
& v_\mathrm{out}(\theta) = \sqrt{ 2 \, \frac{\frac{d F_\mathrm{net}}   {d \omega }}{\frac{ d M_\mathrm{gas}} { d \omega}} \, r_\mathrm{out} } & \nonumber \\ & \approx \sqrt{ 2\left[ a_\mathrm{rad}(r_\mathrm{sub})  - \frac{G \, M_\mathrm{BH}}{r_\mathrm{sub}^2} \right] r_\mathrm{sub}} \left(\frac{r_\mathrm{out}}{r_\mathrm{sub}}\right)^{\frac{1}{2}\gamma - 1} \;.
\label{FullVoutResult}
\end{eqnarray}
Therefore our calculations for the velocity of the gas will have a similarly weak dependence on $r_\mathrm{out}$ as that of the mass-loss rate. 

\subsection{Variation of mass outflow with opening angle}
\label{MdotOpeningAngleResults}

\begin{figure}
\includegraphics[width=0.5\textwidth]{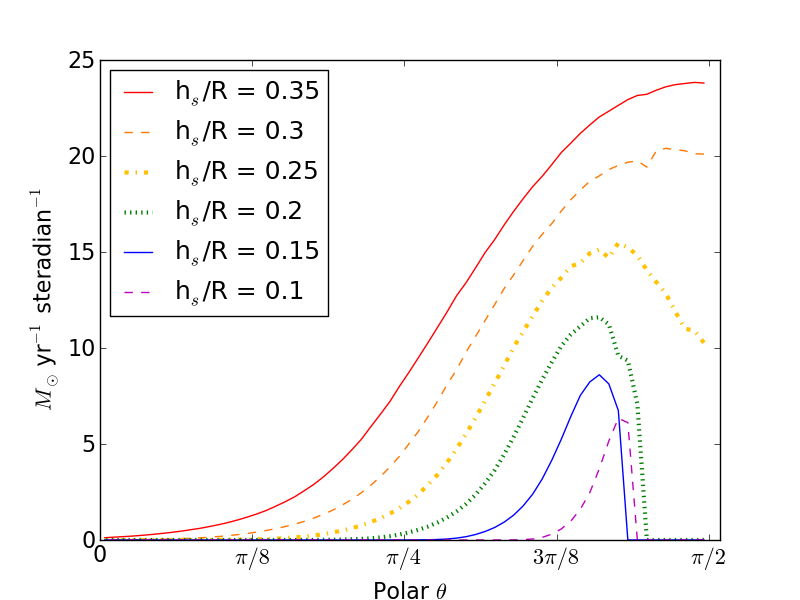}
\caption{Differential mass outflow rate $d\dot{M}/d\omega$ for various opening angles. Parameters correspond to the fiducial values in Table \ref{FiducialParameters}, except for opening angles which vary as indicated (while conserving the total mass in the calculation domain). }
\label{MdotOpeningAngles}
\end{figure}

Figure \ref{MdotOpeningAngles} shows the differential mass outflow rate $d\dot{M}/d\omega$ calculated using equation \ref{MdotDefinition} for disks with various opening angles. The densities of the innermost radial grid zones have been re-scaled to keep the total mass in the calculation domain constant in each case. 
 
The overall mass outflow rate declines with smaller $h_s/R$ due to the increased funneling of radiation into the low-density polar regions. For $h_s/R \geq 0.25$ the differential mass outflow rate peaks at $\theta = \pi/2$, while for $h_s/R \leq 0.25$ the peak is at an intermediate polar angle. This is due to an interplay between the amount of mass available to be cleared away, its inertia, and the gravitational force acting upon it. Near the equator, however, the large amount of gas cannot be unbound by the radiative acceleration and so there is not outflow, even though the force due to radiation is strongest there.  

We emphasize that Figure \ref{MdotOpeningAngles} represents only a snapshot in time of the mass outflow rate for an accreting SMBH. The evolution of $d \dot{M}/ d \omega$ with time is not calculated here and requires a fully coupled radiation-hydrodynamics calculation. Depending on the resulting rearrangement of the gas, the long-term mass loss rate could conceivably be either larger or smaller than the rate calculated here. One possible scenario is that an initially optically thick and puffy disk will blow away gas in the polar region. In the absence of a replenishing mechanism that operates on a timescale shorter than $\overline{t}$, this might cause the disk to become thinner, reducing the tendency for radiation to blow out more gas. Alternatively, the removal of gas from above the near-midplane could change the geometry of the flux sufficiently so as to induce a stronger vertical component, drawing up more gas from the midplane and leading to yet more mass loss. Yet another possibility, already suggested in section \ref{TemperatureAndAccelerationResults}, is that a steady-state inflow/outflow develops with a mass loss rate that hovers close to the instantaneous value computed here. 

\subsection{Variation of mass outflow with other parameters}
\label{MassOutflowOtherParameters}

\begin{figure}
\includegraphics[width=0.5\textwidth]{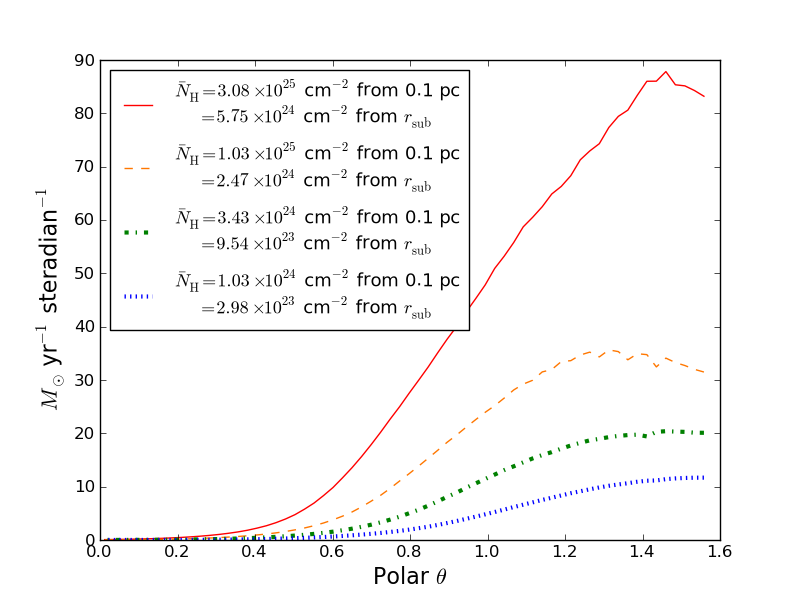}
\caption{Differential mass outflow rate $d\dot{M}/d\omega$ for various sightline-averaged column densities (measured to 0.1 pc from the black hole). Parameters correspond to the fiducial values in Table \ref{FiducialParameters}, except for the average column densities which vary as indicated. The total mass in the calculation domain varies proportionally with the average column density.}
\label{MdotColumnDensities}
\end{figure}

Figure \ref{MdotColumnDensities} shows the differential mass outflow rate $d\dot{M}/d\omega$ for disks with varying column densities (the column density is averaged over all lines of sight, and includes both dusty and non-dusty gas). The variation in column density is directly proportional to variation in the total mass present in the calculation domain. 

As expected, more mass can be ejected when there is more mass present to begin with. However, we find the scaling to be sub-linear: $\dot{M} \propto \overline{N}_\mathrm{H} ^{0.49}$ for the range of column densities included in this study (power-law scaling relations for all the free parameters in the problem will be summarized in section \ref{SummaryScalings}). The slow growth of $\dot{M}$ with column density is due to the fact that at higher column densities, the radiative force on the gas in the densest portions of the disk does not rise as quickly as the mass present there, and so gravity becomes increasingly effective at limiting the outflow rate.

\begin{figure}
\includegraphics[width=0.5\textwidth]{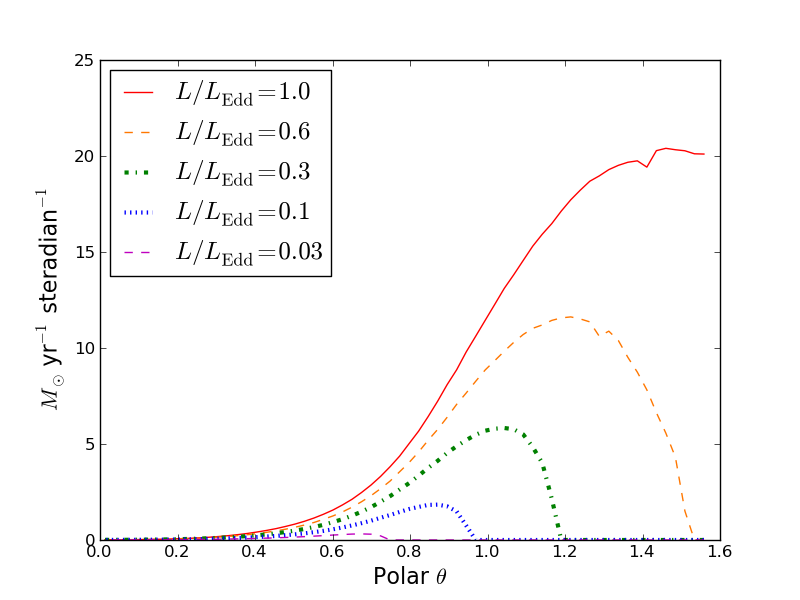}
\caption{Differential mass outflow rate $d\dot{M}/d\omega$ for various luminosities. Parameters correspond to the fiducial values in Table \ref{FiducialParameters}, except for the luminosities which vary as indicated. }
\label{MdotLuminosities}
\end{figure}

Finally, Figure \ref{MdotLuminosities} shows the differential mass outflow rate $d\dot{M}/d\omega$ for calculations with varying black hole luminosities. For higher AGN luminosities, not only is there a higher net force on a column at a given value of $\theta$ for which the net force was already outward (positive), but also the net force becomes positive on columns at larger polar angles. For all other parameters held constant, there exists a critical luminosity at which the radiation force exceeds gravity for all polar angles, and all the gas would be blown away. For opening angle $h_s/R$ = 0.3 and our fiducial mean column density $3.4 \times 10^{24} $ cm$^{-2}$, radial density power-law $\gamma = 1.5$ and black hole mass $M_\mathrm{BH} = 10^8$ $M_\odot$, this critical luminosity is $L / L_\mathrm{Edd} \approx 0.7$. The existence of such a critical luminosity might help to explain the dearth of quasars observed to be radiating at the full value of their inferred Eddington limit, although the precise value of the limiting luminosity presented here should be considered a rough estimate, and may vary with time as the gas rearranges itself following the initial outflow we have estimated.

\subsection{Summary scalings of integrated quantities}
\label{SummaryScalings}

The scalings of $\overline{\tau}_\mathrm{eff}$ with the parameters of the problem, varied one at a time from the fiducial values listed, for a black hole with mass $10^8$ $M_\odot$, can be summarized as follows:
\begin{align}
\overline{\tau}_\mathrm{eff} &= 3.8 \, \left( \frac{h_s/R}{0.3}\right)^{1.5} \nonumber \\
&\times \left( \frac{\overline{N}_\mathrm{H}}{3.4 \times 10^{24} \mathrm{\;cm}^{-2}}\right)^{0.49} \nonumber \\
&\times \left( \frac{L}{1.26 \times 10^{46} \; \mathrm{ergs} \; \mathrm{s}^{-1} }\right)^{-0.13}  \; .
\label{EffectiveTauScalingResults}
\end{align}

For these fits, six data points were used for $h_s/R$ spanning 0.1 to 0.35, five data points were used for $L$ spanning 0.03 to 1, and four data points were used for $\overline{N}_\mathrm{H}$ spanning $10^{24}$ to $3 \times 10^{25}$ cm$^{-2}$. All of these calculations used $\gamma = 1.5$, and the results for changing $\gamma$ generally correlate with the results for the corresponding change in $\overline{N}_\mathrm{H}$.

We may also present a summary scaling relation for the volume-integrated mass outflow rate $\dot{M}$ calculated over the same range of parameters:

\begin{align}
\dot{M} &= 144 \; M_\odot \mathrm{\: yr}^{-1} \, \left( \frac{h_s/R}{0.3}\right)^{2.6} \nonumber \\
&\times \left( \frac{\overline{N}_\mathrm{H}}{3.4 \times 10^{24} \mathrm{\;cm}^{-2}}\right)^{0.62} \left( \frac{L}{1.26 \times 10^{46} \; \mathrm{ergs} \; \mathrm{s}^{-1}}\right)^{1.6} \; . 
\label{MdotScalingResults}
\end{align}

Note that $\overline{N}_\mathrm{H}$ in these scaling relations corresponds to column densities integrated from 0.1 pc to large radii. If instead we use the column density integrated from the edge of the dust sublimation radius outward, then the fiducial column density becomes $9.5 \times 10^{23}$ cm$^{-2}$, the column density power-law in equation \ref{EffectiveTauScalingResults} changes to $0.56$, and the column density power-law in equation \ref{MdotScalingResults} changes to $0.71$. Also note that the rates in equation \ref{MdotScalingResults} correspond to a radius of 32 parsecs from the central SMBH, and there is a weak dependence on radius (no stronger than $r^{1/4}$ when $\gamma = 1.5$; refer to section \ref{MdotScaling}.)

The fiducial value for the mass outflow rate of 144 $M_\odot$ per year may seem surprisingly large. That value was computed for a black hole radiating at its full Eddington luminosity, and at that luminosity the radiative acceleration beats out gravity at all solid angles. Therefore, there is reason to suspect that such a large outflow rate is not sustainable for many gas dynamical times at the parsec scale, as the amount of mass present and the opening angle of the disk readjust during the outflow. The quoted outflow rate also does not incorporate the effects of making the gas distribution clumpy, and as was demonstrated in section \ref{ClumpyResults}, this reduces the integrated force by a factor of $\sim 2$ for significant clumping. Since the mass loss rate should roughly scale as the integrated force to the $1/2$ power (as argued in section \ref{MdotScaling}), the mass outflow rate will be reduced approximately by a factor of $\sqrt{2}$ in the case of significant clumping. On the other hand, the mass-loss rates calculated above correspond to the mass swept up within a radius of approximately 32 parsecs from the central black hole. Extrapolating our results out to 1 kpc, as discussed in section \ref{MdotScalingResults}, might boost the outflow rates by roughly a factor of 2 for $\gamma = 1.5$.

With those caveats in mind, the important conclusions to be drawn from equations \ref{EffectiveTauScalingResults} and \ref{MdotScalingResults} are that the radiation force may reach several times ($\sim 3$) $L /c$, and that the mass outflow rates can easily reach tens of solar masses per year for parameters close to our fiducial values. In the proper circumstances ($\overline{N}_\mathrm{H} \gtrsim 10^{24}$ cm$^{-2}$, $h_s/R \gtrsim 0.25$, and $L/L_\mathrm{Edd} \approx 1$), the mass outflow rates can reach up to 100 $M_\odot$ per year.

It is also interesting to note that the momentum enhancement and mass outflow rate depend relatively strongly on the AGN luminosity and disk opening angle $h_s/R$. The fact that the dependence of the mass outflow rate on luminosity is steeper than the dependence of the radiation force on luminosity may at first seem surprising. This scaling has its origins in an effect noted in section \ref{MassOutflowOtherParameters}, specifically that increasing the luminosity allows the radiation force to exceed gravity for a larger fraction of the solid angle, adding more mass to the outflow than was present at lower luminosities.

Finally, to drive home the point that momentum deposition, not heating, is responsible for the large computed outflow rates, we can estimate the corresponding kinetic luminosities. We use our estimate of the gas velocity as a function of solid angle $v_\mathrm{out}(\theta)$ (equation \ref{VoutDefinition}) to compute the fraction of the accretion luminosity $L$ that goes into kinetic luminosity for the same parameter range used above: 
\begin{align}
& \epsilon_\mathrm{k} \equiv \left[\frac{1}{L}\right]\,\left[ \int{\frac{1}{2}\, \frac{d\dot{M}}{d\omega}(\theta) \, v_\mathrm{out}^2(\theta)\,d\omega }\right] \nonumber \\
& = \left[\frac{2(2 \pi)}{L}\right]\,\left[ \int_0^{\pi/2}{\frac{1}{2}\, \frac{d\dot{M}}{d\omega}(\theta) \, v_\mathrm{out}^2(\theta)\,\sin \theta \, d\theta }\right]  \; .
\end{align}

Once again by varying each parameter one at a time with respect to the fiducial values, our results for a $10^{8}$ solar mass black hole can be summarized as 
\begin{align} 
&\epsilon_\mathrm{k}= 0.009 \, \left( \frac{h_s/R}{0.3}\right)^{1.9} \nonumber \\
&\times \left( \frac{\overline{N}_\mathrm{H}}{3.4 \times 10^{24} \mathrm{\;cm}^{-2}}\right)^{0.19} \times \left( \frac{L}{1.26 \times 10^{46} \; \mathrm{ergs} \; \mathrm{s}^{-1}}\right)^{1.8} \; .
\label{KineticLumScaling}
\end{align}

If the column density is computed by integrating from the dust sublimation radius outward, the corresponding power-law in equation \ref{KineticLumScaling} barely changes at all, increasing to 0.21.

The mass-weighted average velocity of the gas in the outflow will be approximately equal to $\overline{\tau}_\mathrm{eff} L / (c \dot{M})$, although a more accurate value can be obtained by integrating $v_\mathrm{out}(\theta)$ weighted by $dM_\mathrm{gas}/d\theta$ and only counting contributions from solid angles and radii for which gas can be blown out. For our fiducial parameters this average velocity at 32 parsecs from of the computational domain is approximately 1000 km s$^{-1}$.

\subsection{Comparison to Previous Results in the Literature}
It is important to compare and contrast the results of our calculation to previous studies that addressed the same physical problem, including \citet{Pier1992-1}, \citet{Krolik2007} and \citet{Dorodnitsyn2011-1, Dorodnitsyn2012}. The first of those studies included a calculation of steady-state, multi-wavelength radiative transfer through a static dusty torus, but for a different torus geometry and slightly different boundary conditions for the radiative transfer than those used here. By modifying our density distribution to match that of \citet{Pier1992-1}, and preventing dust from sublimating within the pre-determined boundaries of the torus, we find results that are in qualitative agreement with theirs, and quantitatively the values for the components of the radiation force each agree to within 32\% in the central region of the torus (our computed radiation forces are smaller). The primary difference between the studies comes in the range of luminosities considered. \citet{Pier1992-1} point out that for a large range of torus parameters, if $L/L_\mathrm{Edd} \gtrsim 0.1$ the radiation force will overwhelm gravity and lead to an outflow, possibly halting accretion. This conclusion is consistent with our results, although we have sought to quantify the rate of mass outflow and have considered the possibility of simultaneous inflow and outflow processes. 

The work of \citet{Krolik2007} and \citet{Dorodnitsyn2011-1,Dorodnitsyn2012} generate self-consistent density distributions for a torus that is dynamically influenced by radiation pressure, by assuming dynamical equilibrium in the first case and with a numerical radiation-hydrodynamics solver in the second and third. Again, these studies find that disruptive outflows should occur when $L/L_\mathrm{Edd} \gtrsim 0.1$ for Compton-thick torii, consistent with the present work. Furthermore, we find encouraging quantitative agreement with \citet{Dorodnitsyn2012} when we use a gray radiative opacity of $10$ cm$^2$ per gram of gas to match theirs. For a $10^7 M_\odot$ black hole with accretion luminosity of $10^{45}$ erg s$^{-1}$ surrounded by a torus of mass $5 \times 10^4 M_\odot$ enclosed within a radius of 3 parsecs, we find $\dot{M}$ using equations \ref{MdotDefinition}  and \ref{IntegratedMdotDefinition} to be 4.6 $M_\odot$ per year, in agreement with their value of 5 $M_\odot$ per year. If we extend the estimation of $\dot{M}$ to include gas out to a radius of 32 parsecs, then our value of $\dot{M}$ rises to 27 $M_\odot$ per year. Interestingly, when we perform the same calculation with wavelength-dependent radiative opacity, this value drops slightly to 25 $M_\odot$ per year. This decrease is due in part to a larger region of dust sublimation in the wavelength dependent case.

To summarize, our work is consistent with previous studies but seeks to quantify the mass lost in outflows driven by highly luminous ($L / L_\mathrm{Edd} > 0.1$) accretion events, accounting for mass driven away at large radii ($>3$ parsecs). By integrating the momentum deposited in columns of gas, we can make a statement of how the outflows can affect the host galaxy. 

\section{Conclusion}
\label{Conclusion}

We have calculated how radiation pressure from a luminous accretion disk around a SMBH drives a powerful outflow of gas via continuum radiation pressure on dust at distances of 0.1-30 pc from the black hole. Using ambient gas conditions motivated by observational constraints on nuclear obscuration in AGN ($h_s/R \gtrsim 0.25$, $\overline{N}_\mathrm{H} \gtrsim 10^{24}$ cm$^{-2}$) we find that a $10^8 M_\odot$ SMBH radiating at Eddington can drive a wind with velocities of $\sim 1000$'s of km s$^{-1}$ and an instantaneous mass loss rate of $\sim$ 10-100 $M_\odot$ per year (see equation \ref{MdotScalingResults}). For SMBHs with masses $\gtrsim 10^9 M_\odot$, the outflow rates could approach $\sim 1000 M_\odot$ per year.

Radiative heating sublimates the dust out to distances of roughly 0.5 to 1 pc in the mid-plane, and radiation pressure drives away the gas and dust in the polar regions, leaving behind what may constitute the observed dusty torus. The wide-angle bipolarity of these outflows corresponds well to observations of obscured quasars \citep{Greene2012} and Seyfert 2s \citep{Crenshaw2000-2}. Although the radiative acceleration is greatest in the polar regions, the majority of the ejected mass comes from oblique angles where there is a more significant reservoir of gas. By contrast, gas in the equatorial plane is more difficult to unbind because of its large inertia and large integrated gravitational attraction.

The net momentum flux in the resulting outflow can exceed $L / c$ by factors of up to 5 for the parameters studied, as infrared photons interact multiple times during their outward diffusion. As recently demonstrated in the calculations of \citet{Ciotti2010, Novak2011, DeBuhr2011-2}, outflows with these properties have a significant impact on gas in the surrounding host galaxy. Our results for the outflows match reasonably well the observed outflows in local ULIRGs such as Mrk 231 \citep{Rupke2011}. The mass-loss rates and kinetic luminosity fractions we calculate also provide a reasonable match to observations of obscured quasars \citep{Moe2009, Dunn2010}, although our model does not provide a mechanism for launching large amounts of gas at the high velocities ($> 20000$ km s$^{-1}$) observed in these systems at small radii. One possibility is that these quasars are exhibiting both line and continuum radiation pressure driven outflows.

We find that the net effect of the AGN radiation on the surrounding gas is a strong function of the opening angle of the accreting gas at the parsec-scale (the torus). Increasing the opening angle allows more momentum to be deposited in all directions because the mass distribution and emergent radiative flux become more isotropic. We also find a steep dependence of the outflow rate on the luminosity of the accretion disk, because at higher luminosities gas becomes unbound over a greater range of solid angles. This result is also in agreement with the observed anti-correlation between obscured AGN fraction and AGN luminosity \citep{Simpson2005, Hasinger2007, Maiolino2007}, although we are restricting our attention to a single black hole mass. 

Keeping all of our parameters at the fiducial values listed in Table \ref{FiducialParameters} but varying the luminosity, we find that outward radiative acceleration begins to exceed gravity at all angles once $L/L_\mathrm{Edd}$ reaches a value of $\sim 0.7$. This value is subject to uncertainty given our approximate treatment of the gas dynamics, but it may nevertheless help to aid understanding of the relative dearth of broad-line quasars observed to be radiating at their full Eddington luminosity \citep{Kelly2010}.

The effects of dust sublimation play a crucial rule in determining the angular dependence of the radiative force on the torus. The redistribution of flux between polar angles takes place almost entirely in the region of gas in which dust has been sublimated, where infrared radiation is re-emitted from the edge of the dusty gas at angles deviating from the radial direction. Once photons penetrate into the dusty gas, they tend to diffuse radially and deposit momentum almost entirely in the radial direction.

All of the results presented above must be considered in light of the approximations and assumptions that we have used. In particular, while allowing for a range of torus scale heights and masses, we have focused on an initial distribution of gas close to hydrostatic equilibrium, without accounting for the self-consistent dynamical rearrangement of the torus as the outflow takes place, perhaps accompanied by inflow when possible. We have also discounted centrifugal support of the torus, which may lead to an underestimate of the mass loss rate when the integrated radiation force is close to that of gravity. 

Another effect we do not capture in these calculations is the potential for the outflowing gas to develop radiative Rayleigh-Taylor instabilities, which might provide more avenues of photon leakage and reduce the coupling of the radiative momentum to the gas as studied by \citet{Krumholz2012-1}. Concerning this last point, our clumpy gas simulations can provide a preliminary estimate of the extent to which the radiation force would be reduced in the presence of such instabilities. Also, due to the high opacity encountered by the direct ultraviolet radiation from the accretion disk, the direct radiation field plays a much more important role in launching the gas in our calculation than in \citet{Krumholz2012-1}. As demonstrated in \citet{Kuiper2012-1}, the radiative Rayleigh-Taylor instability can be suppressed when the direct radiation field is sufficiently strong.

A fully coupled radiation-hydrodynamic calculation will be needed to fully understand the subsequent behavior of the gas in time. Future work will focus on incorporating the results of this study into hydrodynamic simulations of black hole accretion. In addition to the coupling to the hydrodynamics, more details pertinent to the radiative physics may be addressed in such calculations, including line absorption, anisotropic scattering off of dust, metallicity gradients, and variations in the average dust-to-gas ratio. 

\section*{Acknowledgements}
We thank Jason Dexter, Claude-Andr\'{e} Faucher-Gigu\`{e}re, and Nahum Arav for helpful conversations, along with the referee for thoughtful recommendations. NR is supported by the Department of Energy Office of Science Graduate Fellowship Program (DOE SCGF), made possible in part by the American Recovery and Reinvestment Act of 2009, administered by ORISE-ORAU under contract no. DE-AC05-06OR23100. EQ was supported in part by the David and Lucile Packard Foundation. Support for PFH was provided by NASA through Einstein Postdoctoral Fellowship Award Number PF1-120083 issued by the Chandra X-ray Observatory Center, which is operated by the Smithsonian Astrophysical Observatory for and on behalf of NASA under contract NAS8-03060. This work is supported by the Director, Office of Energy Research, Office of High Energy and Nuclear Physics, and Divisions of Nuclear Physics, of the U.S. Department of Energy under contract No. DE-AC02-05CH11231. This research used resources of the National Energy Research Scientific Computing Center, which is supported by the Office of Science of the U.S. Department of Energy under contract No. DE-AC02-05CH11231.

\end{document}